\documentclass[a4paper,fleqn,usenatbib]{mnras}

\usepackage{newtxtext,newtxmath}

\usepackage[T1]{fontenc}
\usepackage{ae,aecompl}

\usepackage{graphicx}	
\usepackage{amsmath}	
\usepackage{amssymb}	
\usepackage{supertabular}
\usepackage{pdflscape}
\usepackage{ulem}
\usepackage{array}
\usepackage[dvipsnames]{xcolor}

\newcommand{\lta}{\mathrel{\hbox{\raise 0.4 ex \hbox{$<$}\kern
                   -1.5 ex\lower .4 ex\hbox{$\sim$}}}}
\newcommand{\gta}{\mathrel{\hbox{\raise 0.4 ex \hbox{$>$}\kern
                   -1.5 ex\lower .4 ex\hbox{$\sim$}}}}

\title[Statistical analysis of some CP stars]{Statistical analysis of roAp, He-weak and He-rich stars}

\author[S.~Ghazaryan et al.]{
S.~Ghazaryan,$^{1}$\thanks{\fontsize{7.7}{9.2}\selectfont{E-mail: \href{mailto:satenikghazarjan@yahoo.de}{satenikghazarjan@yahoo.de} (SG);
\href{mailto:georges.alecian@obspm.fr}{georges.alecian@obspm.fr} (GA)}}
G.~Alecian$^{2}$\textcolor[rgb]{0,0,1}{\footnotemark[1]} and
A.~A.~Hakobyan$^{1}$
\\
$^{1}$Byurakan Astrophysical Observatory, 0213 Byurakan, Aragatsotn province, Armenia\\
$^{2}$LUTH, CNRS, Observatoire de Paris, PSL University, Universit{\'e} Paris Diderot, 5 Place Jules Janssen, F-92190 Meudon, France\\
}

\date{Accepted 2019 June 15. Received 2019 June 12; in original form 2019 February 15}

\pubyear{2019}

\begin{document}
\label{firstpage}
\pagerange{\pageref{firstpage}--\pageref{lastpage}}
\maketitle

\begin{abstract}
To enlarge our database of Chemically Peculiar (CP) stars, we compiled published data concerning the He-weak and He-rich stars observed by high-resolution spectroscopy techniques during last decades. Twenty He-weak and 28 He-rich stars have been added to the database. We have also distinguished roAp stars from stars previously identified as Ap stars. To deepen our knowledge on statistical overview of the abundance anomalies versus the physical parameters of stars, we compared our data with previous compilations. We applied statistical tests on our data and found interesting correlations for effective temperature and surface gravity for all type of stars and a few correlations for projected rotation velocity only for He-rich stars. Because of the lack of the data we couldn't check whether being a member of binary system is affecting on chemical peculiarities of those stars.
\end{abstract}




\section{Introduction}
\label{sec:intro}

The main goal of our paper is to include in a unique catalogue of chemically peculiar stars rapidly oscillating Ap (roAp) stars, He-weak and He-rich stars, which were observed using high-resolution spectroscopy, and for which detailed abundance determinations were done. He-weak and He-rich stars are now added in the database we have discussed in \citet*[][hereafter referred to as Paper~II]{GhazaryanSAlandHa2018}. We proceed to statistical analysis on them. All these stars are main-sequence chemically peculiar (CP) stars and show peculiar abundances for many elements in their atmospheres.
The abundance peculiarities of CP stars are generally considered to be the consequence of atomic diffusion \citep{MichaudMi1970y} that is a physical process which the effects are detectable if mixing motions are weak enough in outer layers. In that case, the average force on atoms (radiative forces against gravity) leads to a migration of chemical species. Since radiative force is different from one species to the other, abundances become inhomogeneous inside stellar zones where mixing is weak or inexistent. Looking for correlation of abundances with respect to physical parameters of stars, gives important information on processes in play (see the discussion in Section~\ref{sec:evolv}). For instance, a correlation between chemical abundances of a given element with respect to effective temperature suggests that radiative acceleration plays a dominant role for that element, and so, it  may be considered as the strongest signature of atomic diffusion process. If there is a correlation between abundances and surface gravity, this may show a stellar evolution effect on atomic diffusion process. And finally, if there is a correlation between chemical abundances and rotation velocity, one may consider the effect of rotational mixing \citep[see][]{MichaudGeAlGeandRicherJ2015}. These correlations with respect to physical parameters are only indicative, since abundance stratification build-up is a complex non-linear process where the efficiency of atomic diffusion depends on various parameters, and could depend on the evolution history of each star during its life on the main-sequence \citep[see][for evolution of magnetic Bp stars]{BaileyBaLaBa2014}.

RoAp stars are the coolest magnetic Ap(SrCrEu) stars on which rapid pulsations with 5-10~min periods were detected \citep[see][]{KurtzDW1982}. First detection of rapid pulsation with 12.14~min period was in the atmosphere of the Przybylski's star HD101065 by \citet{KurtzDW1978}, and then followed the discovery of pulsations in 5 Ap(SrCrEu) stars. More than 40 roAp stars were discovered before 2008 \citep{BrunttBrNoCuetal2008} in the effective temperature range of 7000K$\,\lta T_{\mathrm{eff}} \lta$\,10000K. In our database only 40 of them are included because of the lack of abundance measurements in other stars identified as belonging to roAp group. The roAp stars are interesting subject for asteroseismology: the detailed study of their pulsations and abundance stratification in their atmosphere will help to argue theoretical models with abundance stratification, which is due to atomic diffusion.

He-weak (or He-w) stars are often found in the range of effective temperature 10000K$\,\lta T_{\mathrm{eff}} \lta$\,20000K. Historically, He-weak stars were identified as a subgroup of CP stars by \citet{NorrisJ1971} because of their significant underabundances of helium and for peculiar abundances of some metals. Even though helium deficiency is a general property of CP stars \citep[see][]{DeutschAJ1947,SargentWLW1964}, it was originally observed more deficient (by the factor 2-15) in He-weak stars than in other CP stars \citep[see][for more details]{MichaudGeAlGeandRicherJ2015}. There are two types of He-weak stars - magnetic and non-magnetic. The non-magnetic He-weak stars are sometimes named P-Ga or phosphorus type stars. They show overabundances of P (up to a factor of 2) and Ga (up to a factor of 5) as compared to their solar values \citep[see][]{LeBlancFr2010}.

The discovery of the first He-rich (or He-r) star $\sigma Ori E$ was done by \citet{BergerJ1956}, and the existence of the group of He-rich star by \citet{OsmerPSandPeterson1974}. He-rich stars are characterised by strong excess of helium (by a factor up to 10). Their effective temperature is generally larger than about 16000K. He-rich stars are mostly slow rotators \citep[99.5 per cent of those stars have $v\,\sin{i} < 130$ km~s$^{-1}$, see][]{ZborilandNorth2000}, they have strong magnetic fields, which are correlated with metal abundances \citep{HungerKGrandHeber1991}. \citet{VauclairVa1975u} explained overabundance of helium in He-rich stars as due to atomic diffusion processes combined with a stellar wind.

In Section~\ref{sec:sample} we represent our updated database and recall how it was developed. In Section~\ref{sec:comp} we show the results of the comparison with previous compilations. In Section~\ref{sec:stat} we look for correlations between fundamental parameters and abundances, as well as ``multiplicity'' (only for single, close binaries) and abundance anomalies. In Section~\ref{sec:discuss}, we discuss the compiled data in the framework of the usual models considered for these stars, and finally, we give some general considerations about the abundance stratification build-up process.

\section{The updated database}
\label{sec:sample}

Our updated database contains additional stars for which abundances were determined by various authors through high resolution spectroscopy. Among them,  28 are identified as He-rich stars, and 20 are He-weak stars. Also, 40 stars that are in the database of Paper~II with the CP-type ApBp are actually roAp stars. They are now identified as roAp. All these new stars, or stars with changed attributes are given in Appendix~\ref{appendix:table} with their physical parameters  as they were provided in publications, including effective temperature, surface gravity, rotation velocity  ($v\,\sin{i}$, actually), and bibliographic references.

As in our previous papers \citet[][hereafter referred to as Paper~I]{GhazaryanGhAl2016} and  Paper~II, we considered the mean of the abundance values of different ions for a given element, and the error-bars were recalculated by the rms standard deviation as in Paper~I and in Paper~II. Here also, to have more homogeneous dataset, if for a given element different abundances by several authors were given, we kept the value from the publication where many other abundances of elements were given. All abundances were rescaled to solar values given by \citet[][hereafter referred to as AGS09]{AsplundMaGrSaSc2009}. The detailed abundances (element per element) for each star are provided as online data. In Appendix~\ref{appendix:table} we give also ``multiplicity'' information (binarity, belonging to a cluster, etc.) of these stars according to Simbad\footnote{\href{http://simbad.u-strasbg.fr/simbad/}{http://simbad.u-strasbg.fr/simbad/}.} archive. In Section~\ref{sec:comp} we present comparison results with previous compilations \citep[see Paper~I, II, and][]{SmithSm1996}.

\section{Comparison with other CP types}
\label{sec:comp}

\begin{figure*}
\centering
\includegraphics[width=18.3cm]{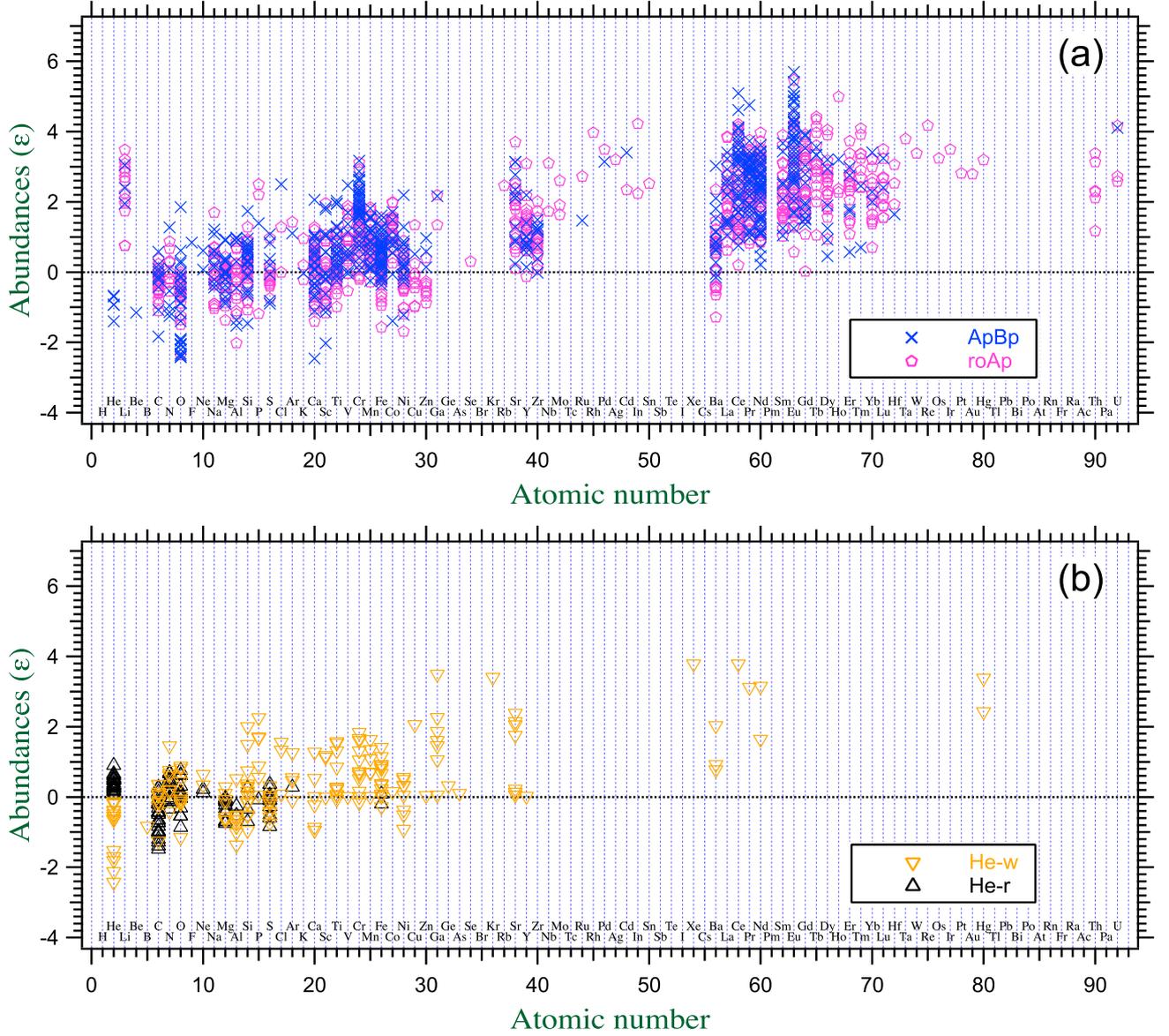}
\caption{Abundances vs. atomic number. Abundances ($\epsilon$) are the logarithm of the abundances divided by the solar AGS09 ones, the zero line corresponds to solar abundances.
(a) Abundances for roAp (open pentagons) and cool ApBp stars (crosses). Abundances of cool ApBp stars are taken from Paper~II.
(b) Abundances for He-weak (top-down triangles) and He-rich stars (top-up triangles).}
\label{fig:layout_panelshe}
\end{figure*}

In Fig.~\ref{fig:layout_panelshe}a we compare the abundance peculiarities found in roAp stars to those of cool ApBp stars (discussed in Paper~II) that are in the same effective temperature range ($T_{\mathrm{eff}} \le$10300\,K, see Sec.\,\ref{sec:roap}), and for which no oscillations were presently found. See discussion about roAp stars in Sec. \ref{sec:roap}.

As shown in Fig.~\ref{fig:layout_panelshe}a, there are no helium and neon abundances determinations for roAp stars. Oxygen is underabundant in both type stars, but for ApBp stars there are a few measurements of overabundance which is up to about 100. The same situation holds for titanium. Iron underabundance is more pronounced by one order of magnitude in roAp stars, (this is related to their low effective temperature, see Section~\ref{sec:roap}).
It is important to note that among these differences there are several metal abundance determinations that are absent for  ApBp stars.

For He-weak stars, we compare our compilation with Smith's one \citep{SmithSm1996}. Considering his fig.~5, we notice that in our compilation helium is deficient by a factor of about 2.5, while in Smith's data it is a bit stronger (see Fig.~\ref{fig:layout_panelshe}b). On the contrary, phosphorus excess is a bit more (300 instead of 100), we compiled only one krypton abundance value (for HD~120709), which excess factor reaches 2500. The comparison results obtained for gallium, sulphur, silicon, manganese and iron, are more or less the same. Notice that HD149363 is very atypical, since its effective temperature is $30000$K according to \citet{ZborilandNorth2000}, all others He-w stars being cooler than $T_{\mathrm{eff}} \approx$\,18500K. Considering the trend of metals overabundances, it appears to be  a little bit less pronounced but more or less similar to what is found for ApBp stars (including HgMn).

Here also we confirm that in Sr-Ti He-weak stars carbon and silicon abundances are normal, whereas titanium, chromium and iron are enhanced \citep[see][]{VilhuOTuandBo1976}. We found more rare earth elements enhanced abundances, which are absent in Smith's compilation, but for detailed study of them more values are needed.

He-rich stars, which show age-dependent increase of helium abundance \citep[see][]{ZborilGlandNorth1994}, exhibit often light elements deficiencies such as carbon, nitrogen and oxygen, and small deficiency of metals such as magnesium, aluminium, silicon, and also deficiency of sulphur. Neon, argon and iron abundances are close to solar ones (Fig.~\ref{fig:layout_panelshe}b).

\section{Statistical analysis}
\label{sec:stat}

It is interesting to know if the abundances in the atmospheres of all CP stars (that may be understood in the framework of atomic diffusion theory) is somehow correlated with physical parameters of the star such as  effective temperature, surface gravity and rotation velocity. To check the possible correlations, following Paper~II we applied the Spearman's rank correlation test\footnote{Spearman's rank test performs a hypothesis test on a pair of two variables with null hypothesis that they are independent, and alternative hypothesis that they are not. Usually, one accepts the alternative hypothesis of the test when $p$-value is less than 5 per cent \citep[see][for more details]{FeigelsonBabu2012}. Spearman's coefficient $\rho$ is a nonparametric measure of rank correlation ($\rho\in[-1;1]$), it assesses how well the relationship between two variables can be described using a monotonic function.} \citep{Spearman1904} between abundances and fundamental parameters. Our results of the correlation test for roAp, He-weak and He-rich stars are presented in Table~\ref{table:rankscp}. All significant correlations ($p$-value $\le 0.05$), which were found for the mentioned CP types, are marked in boldface in Table~\ref{table:rankscp}, and all marginal cases ($0.05<$$p$-value $< 0.06$) are marked as underlined. As in Paper~II, we think that small number statistics for elements with less measurements make the test ineffective, this is why we selected a threshold of 11 measurements and applied the test only for each element measured in more than that threshold value (for roAp and He-abnormal stars).

Our preliminary statistical results show that for roAp stars there is a correlation between oxygen, silicon, calcium, titanium, chromium, manganese, iron, nickel, lanthanum, terbium, ytterbium, and $T_{\rm{eff}}$, marginal correlation exists for sodium and $T_{\rm{eff}}$.

There is also a correlation between lithium, scandium, vanadium, europium, dysprosium and surface gravity. We couldn't find any correlation between an element abundance and 	projected rotation velocity in roAp stars.

For He-weak stars we found correlation between iron abundance and $T_{\rm{eff}}$. There is no significant correlation between abundances and surface gravity, as well as abundances and projected rotation velocity.

For He-rich stars, correlation between helium, nitrogen abundances and effective temperature was detected. For surface gravity, we found correlation for helium, carbon, nitrogen and oxygen abundances, and for projected rotation velocity, we found correlations for oxygen and magnesium abundances, and marginally for helium abundance.

It is important to note that the abundances and physical parameters of stars, which were measured by different authors with different techniques, could affect the statistical results. However, for several stars in our database with given parameters and obtained abundances from different sources, the differences in determinations are always within the errors of any unique measurement, so the impact of inhomogeneous data on our statistical results should be negligible.

To check whether multiplicity is playing a role in the abundance anomalies, we have considered single CP stars (in this case {\it{multiplicity}} is equal to 1) and those being in binary systems (in this case {\it{multiplicity}} is equal to 2)\footnote{It would be more reasonable to consider only the spectroscopic/interacting binaries (multiplicity is equal to 4) but because of the lack of published observational data, particularly for those binaries, we cannot perform such analysis.}. Following Paper~II, we have applied the well-known {\it{Anderson-Darling}} (AD) test\footnote{The null hypothesis for the two-sample nonparametric AD test corresponds to the case when two distributions are drawn from the same parent population, and the alternative hypothesis that they are not (again, with the threshold of 5 per cent for $p$-values). For more details, the reader is referred to \cite{EngmannCousineau11}.} for roAp, He-weak and He-rich stars. With this test, we do not find any relation between abundance anomalies and multiplicity in all mentioned CP type stars, possibly because of the lack of data. However, our results on the multiplicity test should be taken into consideration with caution, because of Simbad data incompleteness.

\begin{table}
\centering
\caption{Spearman's rank test results for roAp, He-weak (He\_w) and He-rich (He\_r) stars.
         Statistically significant correlations are shown in boldface ($p$-value $\le 0.05$),
         marginal ones are underlined ($0.05<$$p$-value $< 0.06$).}
\tabcolsep 3.3pt
\label{table:rankscp}
\begin{tabular}{|c|c|c|c|c|c|c|c|c|c|}
\hline
&\multicolumn{3}{c|}{$\epsilon(T_{\rm{eff}})$}&\multicolumn{3}{c|}{$\epsilon(\log{g})$}&\multicolumn{3}{c|}{$\epsilon(v\,\sin{i})$}\\
\hline
$Elements$&$\rho$&$p$&$N$&$\rho$&$p$&$N$&$\rho$&$p$&$N$\\
\hline
&\multicolumn{9}{c|}{$roAp$}\\
\hline
Li & 0.23 & 0.452 & 13 & \textbf{0.61} & \textbf{0.027} & \textbf{13} & -0.32 & 0.286 & 13\\
O & \textbf{-0.58} & \textbf{0.005} & \textbf{22} & 0.23 & 0.299 & 22 & -0.17 & 0.449 & 22\\
Na & \underline{0.44} & \underline{0.053} & 20 & -0.14 & 0.551 & 20 & 0.14 & 0.557 & 20\\
Si & \textbf{0.36} & \textbf{0.043} & \textbf{32} & 0.02 & 0.908 & 32 & 0.02 & 0.931 & 32\\
Ca & \textbf{0.65} & \textbf{0.000} & \textbf{33} & -0.01 & 0.976 & 33 & 0.11 & 0.545 & 33\\
Sc & 0.20 & 0.377 & 22 & \textbf{0.42} & \textbf{0.049} & \textbf{22} & -0.22 & 0.321 & 22\\
Ti & \textbf{0.73} & \textbf{0.000} & \textbf{28} & 0.19 & 0.331 & 28 & -0.13 & 0.495 & 28\\
V & -0.19 & 0.409 & 21 & \textbf{0.54} & \textbf{0.012} & \textbf{21} & -0.01 & 0.969 & 21\\
Cr & \textbf{0.81} & \textbf{0.000} & \textbf{36} & -0.27 & 0.114 & 36 & 0.19 & 0.276 & 36\\
Mn & \textbf{0.59} & \textbf{0.003} & \textbf{24} & -0.11 & 0.614 & 24 & -0.09 & 0.677 & 24\\
Fe & \textbf{0.82} & \textbf{0.000} & \textbf{36} & -0.06 & 0.729 & 36 & -0.03 & 0.864 & 36\\
Ni & \textbf{0.76} & \textbf{0.000} & \textbf{24} & 0.27 & 0.197 & 24 & 0.092 & 0.669 & 24\\
La & \textbf{0.46} & \textbf{0.020} & \textbf{25} & 0.14 & 0.499 & 25 & 0.10 & 0.632 & 25\\
Eu & 0.24 & 0.174 & 33 & \textbf{0.35} & \textbf{0.048} & \textbf{33} & -0.30 & 0.095 & 33\\
Tb & \textbf{-0.53} & \textbf{0.036} & \textbf{16} & -0.21 & 0.431 & 16 & -0.18 & 0.493 & 16\\
Dy & -0.10 & 0.698 & 18 & \textbf{-0.56} & \textbf{0.015} & \textbf{18} & -0.43 & 0.075 & 18\\
Yb & \textbf{0.62} & \textbf{0.014} & \textbf{15} & -0.29 & 0.289 & 15 & 0.02 & 0.955 & 15\\
\hline
&\multicolumn{9}{c|}{$He\_w$}\\
\hline
Fe & \textbf{0.55} & \textbf{0.043} & \textbf{14} & 0.19 & 0.525 & 14 & -0.26 & 0.399 & 13\\
\hline
&\multicolumn{9}{c|}{$He\_r$}\\
\hline
He & \textbf{0.46} & \textbf{0.019} & \textbf{26} & \textbf{-0.40} & \textbf{0.043} & \textbf{26} & \underline{-0.44} & \underline{0.051} & \underline{20}\\
C & -0.29 & 0.138 & 27 & \textbf{-0.44} & \textbf{0.022} & \textbf{27} & -0.02 & 0.923 & 22\\
N & \textbf{-0.41} & \textbf{0.038} & \textbf{26} & \textbf{0.39} & \textbf{0.048} & \textbf{26} & 0.23 & 0.311 & 22\\
O & -0.12 & 0.635 & 18 & \textbf{0.56} & \textbf{0.016} & \textbf{18} & \textbf{0.55} & \textbf{0.034} & \textbf{15}\\
Mg & -0.23 & 0.358 & 18 & 0.32 & 0.201 & 18 & \textbf{0.63} & \textbf{0.011} & \textbf{15}\\
\hline

\end{tabular}
\end{table}

\section{Discussion}
\label{sec:discuss}

\subsection{roAp stars}
\label{sec:roap}

RoAp stars are essentially defined as being Ap stars for which rapid oscillations are detected. At first view, their abundance peculiarities look like those of non-oscillating Ap stars if one excepts the very peculiar case of Przybylski's star (HD101065). Pulsations are supposed to be excited by $\kappa$-mechanism \citep[see for instance][]{BalmforthBaCuDoetal2001}, which suggests a strong dependence on opacities in the atmosphere, and so, on the abundance stratifications produced by atomic diffusion.

\begin{figure}
\includegraphics[width=9cm]{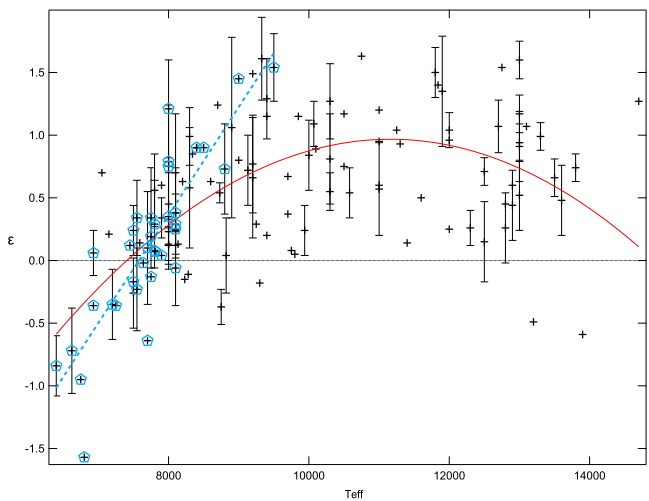}
\caption{Fe abundances ($\epsilon$) in ApBp and roAp stars vs. $T_{\rm{eff}}$. The red solid curve is a polynomial fit showing the correlation of $\epsilon$ and $T_{\rm{eff}}$ for all ApBp stars (including roAp) of our database for which iron abundance was measured. These Fe abundances and the red solid curve were shown in fig.~5 of Paper~II. The light-blue symbols surround those of that stars that are identified as roAp stars in the present work. The light blue dashed line is the linear fit for these stars.}
\label{fig:fe_apbp_roap}
\end{figure}

To be more specific in analyzing observational differences between the whole ApBp group and the subgroup of roAp stars, we have reconsidered in Fig.~\ref{fig:fe_apbp_roap} the fig.~5 of Paper~II, and revealed the contribution of roAp stars. Fig.~\ref{fig:fe_apbp_roap} shows the iron abundances (with respect to the Sun) vs $T_{\rm{eff}}$ for all ApBp stars (black crosses) including roAp stars, and the corresponding polynomial fit (red line) shown in Paper~II \citep[such a trend may found also in][]{BaileyBaLaBa2014}. The light blue symbols surround those of stars that are now identified as roAp stars in our database. We have checked that the $T_{\rm{eff}}$ of the maximum of the polynomial fit (red line) does not change significantly (nor the shape) when the fit is recomputed removing the roAp stars, it remains close to 11000K, slightly shifted toward 11300K. However we notice that roAp stars have a major (or even exclusive) contribution at the cool end of the ApBp group, and are clearly responsible for the Fe deficiency we noticed in Section~\ref{sec:comp}.

\begin{figure}
\begin{center}$
\begin{array}{@{\hspace{0mm}}c@{\hspace{0mm}}}
\includegraphics[width=0.95\hsize]{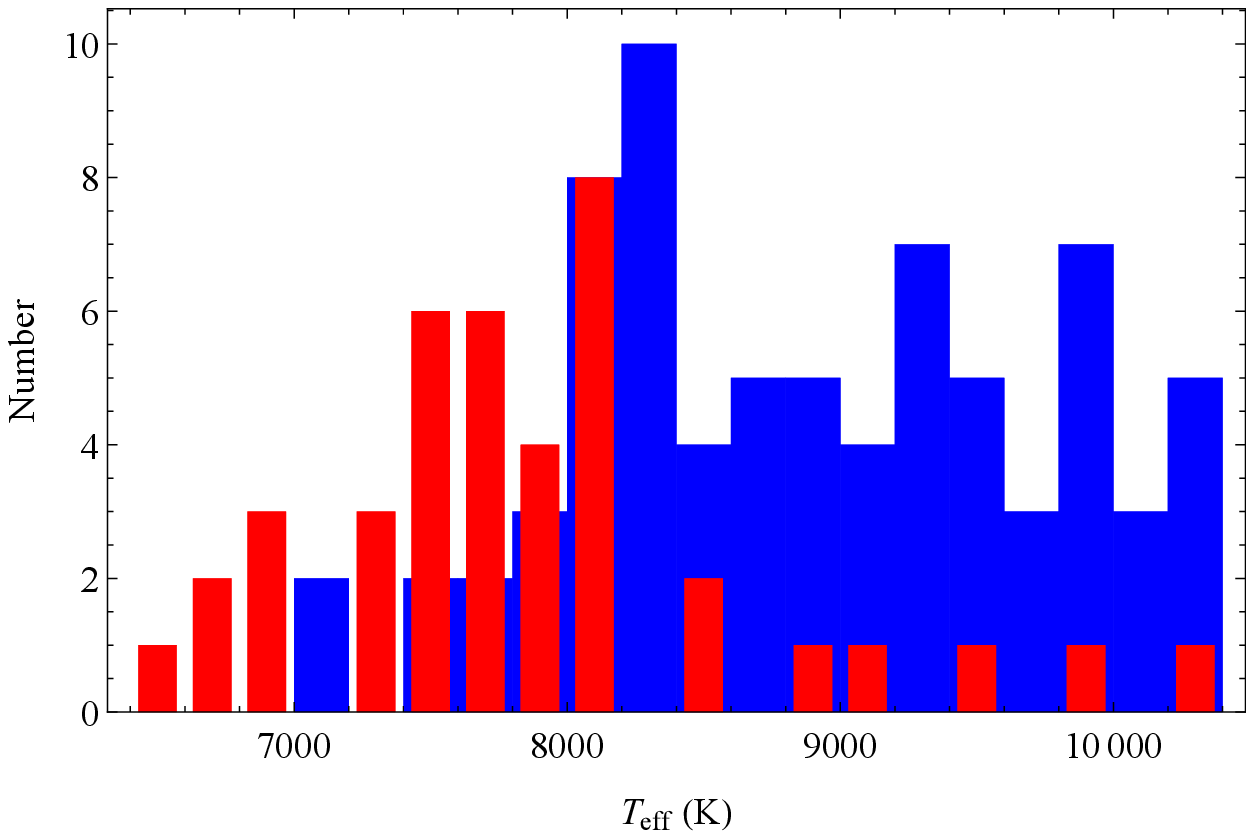}\\
\includegraphics[width=0.95\hsize]{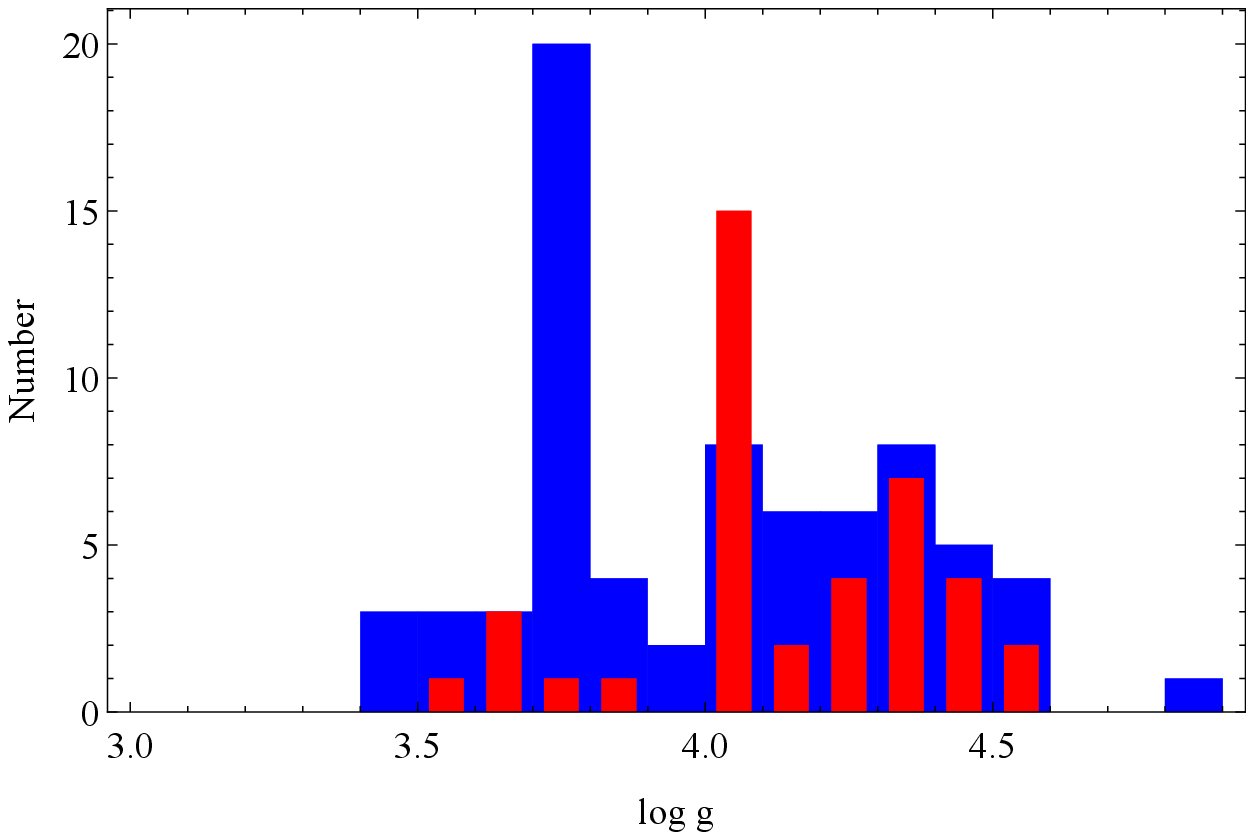}\\
\includegraphics[width=0.95\hsize]{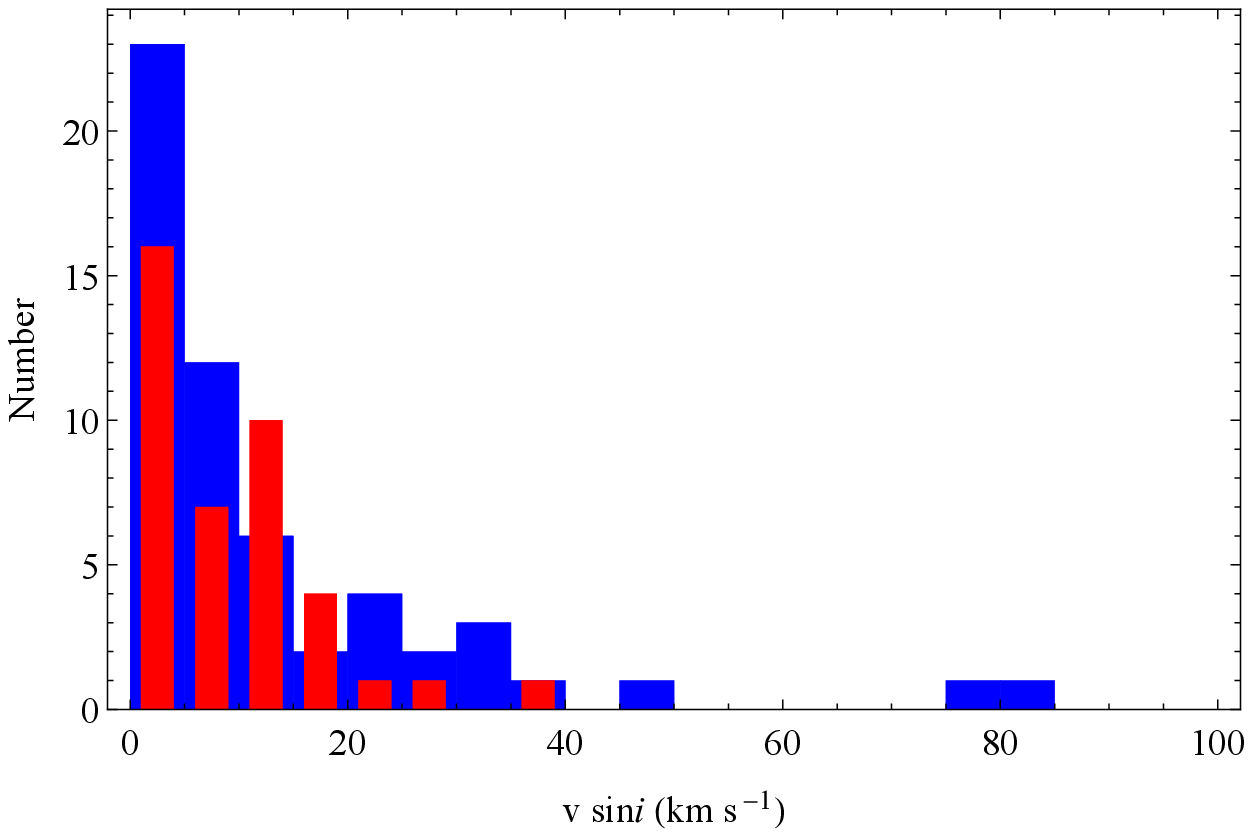}
\end{array}$
\end{center}
\caption{Histograms of the $T_{\rm{eff}}$, $\log{g}$, $v\,\sin{i}$ for roAp and cool ApBp stars by red and blue colours, respectively.
HD320764 that has 225 $\rm{km~s}^{-1}$ is outside of the $v\,\sin{i}$ histogram).}
\label{fig:layouthisto_roap_ap}
\end{figure}

We show in Fig.~\ref{fig:layouthisto_roap_ap}, the histograms of the $T_{\rm{eff}}$, $\log{g}$, and $v\,\sin{i}$ for roAp stars, to be compared to those for cool ApBp stars where rapid oscillation was not detected. The hottest roAp in our database (HD9996a\footnote{Notice that the abundances for elements lighter than Z=46 are not available for this star except for vanadium.}) has $T_{\rm{eff}}=10300$K, therefore this is the upper temperature limit we impose to our selection of cool ApBp stars (hereafter, we will call them simply Ap stars). We notice that for the roAps the histogram of $T_{\rm{eff}}$ has a maximum around 8000K followed by a sharp decrease. There is also a maximum close to the same temperature for Ap stars, however, contrarily to roAp stars, it is not followed by a steady decrease. Considering $\log{g}$, the maximum is close to 4.0 for roAp stars, while it appears smaller for Ap stars that suggests that roAp stars are less evolved. Concerning the $v\,\sin{i}$, there is no roAp stars with higher projected velocity than 35 $\rm{km~s}^{-1}$, contrarily to Ap stars where higher velocities are found. We applied AD test on roAp and cool ApBp stars' physical parameters ($T_{\rm{eff}}$, $\log{g}$, and $v\,\sin{i}$) and in two cases (for $T_{\rm{eff}}$ and $\log{g}$) the distributions of physical parameters for roAp and cool ApBp stars are significantly different ($p$-value $< 0.05$). For $v\,\sin{i}$, the $p$-value $=0.357$, which means that the distributions are consistent with each other.

\subsection{He-weak stars}
\label{sec:Hew}

Since one has only 20 He-w stars in our database, and since determinations of He abundance from spectral data (that determines the belonging to this CP group) may be especially difficult, one cannot safely proceed with histograms as we do above for ApBp stars.

In Section~\ref{sec:comp}, we have noticed that the trend of metal overabundances is more or less similar to what is found for ApBp stars (including HgMn). This is not very surprising since CP stars with $T_{\mathrm{eff}} \lta$\,18000K should be generally He deficient according to atomic diffusion model \citep[see][]{MichaudGeAlGeandRicherJ2015}. Actually, one could question the existence of a distinct He-w type. One cannot exclude that these stars simply belong to the other CP types with $T_{\mathrm{eff}} \lta$\,18000K.

\subsection{He-rich stars}
\label{sec:Her}

Considering the histograms for the 28 He-r stars we have included in the database, we do not see very noteworthy structure in the distribution of the  three stellar parameters. Apart the hottest member of the group (HD68450), all He-r have $16000 \lta T_{\mathrm{eff}}(\rm{K}) \lta 26000$, with a moderate peak around 23000K. Their $\log{g}$ most often (but not always) are larger than 4.0, which suggests that they are less evolved than usual CP type stars.

He-rich stars may be understood in the framework of the unified model proposed initially by \citet{VauclairVa1975u}, and recently discussed in  \citet{MichaudGeAlGeandRicherJ2015} (see especially Section~8.3.2.3). In a few words, the model considers that atomic diffusion has to be combined with the increasing mass-loss (or wind) with increasing effective temperature. Elements that undergo strong radiative acceleration and which appear overabundant in ApBp (and HgMn) stars, are more easily expelled from the atmosphere of He-rich stars (that are hotter). This is because, the wind and the diffusion velocities are both positive, and wind velocity is expected to be larger than for ApBp stars. This doesn't mean that these elements will become underabundant, since mass-loss corresponds to a continual flux of matter from deep layers where abundances may be homogenously distributed. Stronger is the mass loss rate, faster is the replacement of the superficial matter by the deeper one. So, a strong mass-loss imped atomic diffusion in stratifying abundances. Elements with weaker radiative accelerations (like He, or CNO), which sink by gravitational settling in usual CP stars (negative diffusion velocity), may accumulate in the atmosphere of He-r stars, according to the value of the wind velocity compared to the one of diffusion. Notice that this simple model is not generally accepted \citep{KrtickaKrKuGr2006}, but as much as we know, no alternate model is  presently proposed. One cannot say that abundances in He-r stars other than the ones for He, as shown in Fig.~\ref{fig:layout_panelshe}b, support strongly the simple unified model that is presently still qualitative. However, the underabundances observed for CNO and some heavier metals appear less pronounced than in ApBp stars as could be expected in the framework of the unified model. We would like to emphasize that numerical models including atomic diffusion together with mass-loss are presently lacking for hot CP stars (namely He-r).

\subsection{Abundance stratification build-up vs. stellar parameters and evolution}
\label{sec:evolv}

In this work, as in our previous Papers~I and II, we have looked for correlations of superficial abundances with respect to physical parameters of stars. This is motivated by the fact that atomic diffusion cannot be directly observed, but only through its secondary effects such as for atmospheres superficial abundance anomalies or element distribution, for interiors through opacity effects (convection, pulsation, for instance). Because any star is different from each other, and because atomic diffusion is extremely sensitive to any perturbation, theoretical models at this time may only address general trends, this is why looking for such correlations is helpful. The reader may find in \citet{AlecianAl2015} a list of physical processes that should be considered for atmospheres (some of them are not yet included in numerical models simulating the abundance stratification build-up).

As a good example of the interest in using observed abundance correlation with effective temperature, it is common to cite the trend of Mn overabundances found by \citet{SmithSmDw1993r} for HgMn stars. It was previously predicted by the theoretical study of manganese NLTE radiative acceleration by \citet{AlecianAlMi1981}. This trend has been identified as to be due to Mn stratification process in optically thin layers, above $\tau\approx 1$, where diffusion time scales are much shorter than stellar evolution time as shown by \citet{AlecianAlStDo2011}. These authors have also shown that abundance stratification in atmospheres can rapidly adjust to changes in atmospheric parameters (in less than 10 to 1000 years according to the element). Therefore, the trend found for Mn is not due to secular evolution: even if the effective temperature changes with the star age, the spread of $T_{\rm{eff}}$ is mainly due to the variety of stellar masses for which stratification of abundances can occurs. However, this doesn't mean that secular evolution as found by \citet{BaileyBaLaBa2014} do not affect the observed abundance peculiarities. For instance, considering the atmospheric abundance evolution with time observed by \citet{BaileyBaLaBa2014} for many elements, one may expect that the shape of abundance peculiarity trends should be different comparing stellar clusters having different ages. On another hand for instance, one expects that a young HgMn stars with $T_{\rm{eff}}$ slightly hotter than 10000K, may become Am stars during their evolution on the main sequence \citep[as suggested by][Section~9.1.1]{MichaudGeAlGeandRicherJ2015}\footnote{During evolution   $T_{\rm{eff}}$ decreases, and that may trigger the appearance of a superficial convection zone which characterizes the structure of AmFm stars. HgMn stars are considered to be the continuation of the AmFm group to hotter effective temperatures. Both groups are non-magnetic.}. The situation is different when one considers the gravity, which spreads from 3.4 to 4.6 in dex. Because these different $\log{g}$ are due to stellar evolution on the main-sequence. This is what we mean by writing in Section~\ref{sec:intro} that correlation with gravity may show an effect of evolution.
Most of published discussions confronting theoretical models for atmospheres and observations have been done considering \emph{equilibrium solutions} of equations for abundance stratifications\footnote{These solutions assume that stratification build-up reaches a state such that atomic diffusion fluxes are zero everywhere in the atmosphere \citep[see detailed discussion in][]{StiftStAl2016}.}. However, as did also in \citet{BaileyBaLaBa2014}, we draw the reader's attention on the limits of this approach\footnote{To consider equilibrium solution was justified, because stationary solutions for atmospheres  may be obtained only through very heavy numerical calculations that are achieved only very recently by solving the time-dependent continuity equation for concentrations in optically thin media.}. Indeed, equilibrium solution for a given element corresponds to the maximum abundance that can be supported by the radiation field in each layer. The stratification build-up is actually complex and several numerical simulations carried out by \citet{StiftStAl2016}, and more recently by \citet{AlecianAlSt2019} have shown that stratification build-up converge rather to stationary solutions quite different from equilibrium ones, especially when a stellar mass loss (or a wind) is assumed. In addition, one has to consider for ApBp stars the existence of a strong magnetic field that always affects strongly atomic diffusion velocities (as well by its strength than by its orientation). Observations of \citet{BaileyBaLaBa2014} show quite well that abundance peculiarities are correlated with the magnetic field intensity. Notice that the first 3D simulations in magnetic atmospheres have been carried out by \citet{AlecianAlSt2017} in equilibrium hypothesis (same studies for non-equilibrium hypotheses are in progress). It is clear that correlation with stellar parameters as shown in the present work and others, cannot reveal this complex reality, except through the well known dispersion of abundances from star to star that may be easily noticed in Fig.~\ref{fig:layout_panelshe} or \ref{fig:fe_apbp_roap}.

\section{Conclusions}
\label{sec:conc}

In this work we present adjunction to our database (discussed in Paper~II) of 88 CP stars with their fundamental parameters and chemical abundances. As in Paper~II, these data have been obtained by compiling the data obtained through high resolution spectroscopy and published by various authors. The adjunction consists in 20 He-weak, and 28 He-rich stars. We have also updated the CP-type of 40 ApBp stars that are actually roAp stars. This new compilation allows us to compare the abundance peculiarities observed for these CP-types to those of the types considered in Paper~II.

For each CP-type, we applied Spearman's rank correlation test on the chemical abundances according to physical parameters (effective temperature, surface gravity and projected rotation velocity). We have found that there are significant correlations for effective temperature and surface gravity and a few ones with rotation velocity in He-rich stars. As we mentioned in Paper~II, all these correlations could help in identifying which of the physical parameters could have a significant effect on abundance stratification process in these stars. Among the limits of such a study, we would like to point out that an important source of errors on abundances published in the literature is the fact that abundance determinations are generally done assuming homogeneous distribution of elements in CP stars atmospheres, although CP stars (except for AmFm or possibly He-r stars) have certainly strongly stratified abundances in their atmospheres and inhomogeneous horizontal distributions of elements in magnetic cases. Our statistical results are consistent with the important role of radiative accelerations in roAp and He-weak stars. We may also question the existence of a distinct He-weak CP-type. He-rich stars appear to be clearly different from other main-sequence CP-types stars. This could be possibly due, as proposed by the unified model \citep{VauclairVa1975u}, to the competition between a strong mass-loss (increasing with $T_{\rm{eff}}$) with atomic diffusion. However this unified model still needs to be confirmed by numerical simulations.

In addition, we tried to apply AD test to only single and binary type stars but possibly because of the lack of data we couldn't get any relation between abundance anomalies and multiplicity.
We are convinced that more stars in the database should present better results about correlations. Therefore, large catalogs such as the one of Gaia mission will have a significant impact in near future.

\section*{Acknowledgements}

We would like to thank the referees for excellent comments that improved the clarity of this paper. This work was supported by the RA MES State Committee of Science, in the frames of the research project N\textsuperscript{\underline{o}} 16YR--1C034.




\bibliographystyle{mnras}
\bibliography{some_cp_stars}

\begin{thebibliography}{}
\makeatletter
\relax
\def\mn@urlcharsother{\let\do\@makeother \do\$\do\&\do\#\do\^\do\_\do\%\do\~}
\def\mn@doi{\begingroup\mn@urlcharsother \@ifnextchar [ {\mn@doi@}
  {\mn@doi@[]}}
\def\mn@doi@[#1]#2{\def\@tempa{#1}\ifx\@tempa\@empty \href
  {http://dx.doi.org/#2} {doi:#2}\else \href {http://dx.doi.org/#2} {#1}\fi
  \endgroup}
\def\mn@eprint#1#2{\mn@eprint@#1:#2::\@nil}
\def\mn@eprint@arXiv#1{\href {http://arxiv.org/abs/#1} {{\tt arXiv:#1}}}
\def\mn@eprint@dblp#1{\href {http://dblp.uni-trier.de/rec/bibtex/#1.xml}
  {dblp:#1}}
\def\mn@eprint@#1:#2:#3:#4\@nil{\def\@tempa {#1}\def\@tempb {#2}\def\@tempc
  {#3}\ifx \@tempc \@empty \let \@tempc \@tempb \let \@tempb \@tempa \fi \ifx
  \@tempb \@empty \def\@tempb {arXiv}\fi \@ifundefined
  {mn@eprint@\@tempb}{\@tempb:\@tempc}{\expandafter \expandafter \csname
  mn@eprint@\@tempb\endcsname \expandafter{\@tempc}}}

\bibitem[\protect\citeauthoryear{{Adelman}, {Caliskan}, {Cay}, {Kocer}  \&
  {Tektanali}}{{Adelman} et~al.}{1999}]{AdelmanAdCaCaetal1999}
{Adelman} S.~J.,  {Caliskan} H.,  {Cay} T.,  {Kocer} D.,   {Tektanali} H.~G.,
  1999, \mn@doi [\mnras] {10.1046/j.1365-8711.1999.02435.x}, \href
  {http://adsabs.harvard.edu/abs/1999MNRAS.305..591A} {305, 591}

\bibitem[\protect\citeauthoryear{{Alecian}}{{Alecian}}{2015}]{AlecianAl2015}
{Alecian} G.,  2015, \mn@doi [MNRAS] {10.1093/mnras/stv2205}, \href
  {http://adsabs.harvard.edu/abs/2015MNRAS.454.3143A} {454, 3143}

\bibitem[\protect\citeauthoryear{{Alecian} \& {Michaud}}{{Alecian} \&
  {Michaud}}{1981}]{AlecianAlMi1981}
{Alecian} G.,  {Michaud} G.,  1981, \mn@doi [ApJ] {10.1086/158803}, \href
  {http://adsabs.harvard.edu/abs/1981ApJ...245..226A} {245, 226}

\bibitem[\protect\citeauthoryear{{Alecian} \& {Stift}}{{Alecian} \&
  {Stift}}{2017}]{AlecianAlSt2017}
{Alecian} G.,  {Stift} M.~J.,  2017, MNRAS, \href
  {http://adsabs.harvard.edu/abs/2017arXiv170208322A} {468, 1023}

\bibitem[\protect\citeauthoryear{Alecian \& Stift}{Alecian \&
  Stift}{2019}]{AlecianAlSt2019}
Alecian G.,  Stift M.~J.,  2019, \mn@doi [\mnras] {10.1093/mnras/sty3003}, 482,
  4519

\bibitem[\protect\citeauthoryear{{Alecian}, {Stift}  \& {Dorfi}}{{Alecian}
  et~al.}{2011}]{AlecianAlStDo2011}
{Alecian} G.,  {Stift} M.~J.,   {Dorfi} E.~A.,  2011, MNRAS, 418, 986

\bibitem[\protect\citeauthoryear{{Alentiev}, {Kochukhov}, {Ryabchikova},
  {Cunha}, {Tsymbal}  \& {Weiss}}{{Alentiev}
  et~al.}{2012}]{AlentievAlKoRyetal2012}
{Alentiev} D.,  {Kochukhov} O.,  {Ryabchikova} T.,  {Cunha} M.,  {Tsymbal} V.,
   {Weiss} W.,  2012, \mn@doi [\mnras] {10.1111/j.1745-3933.2011.01211.x},
  \href {http://adsabs.harvard.edu/abs/2012MNRAS.421L..82A} {421, L82}

\bibitem[\protect\citeauthoryear{{Allende Prieto} \& {Lambert}}{{Allende
  Prieto} \& {Lambert}}{1999}]{Allende-PrietoAlLa1999}
{Allende Prieto} C.,  {Lambert} D.~L.,  1999, \aap, \href
  {http://adsabs.harvard.edu/abs/1999A%26A...352..555A} {352, 555}

\bibitem[\protect\citeauthoryear{{Asplund}, {Grevesse}, {Sauval}  \&
  {Scott}}{{Asplund} et~al.}{2009}]{AsplundMaGrSaSc2009}
{Asplund} M.,  {Grevesse} N.,  {Sauval} A.~J.,   {Scott} P.,  2009, \mn@doi
  [\araa] {10.1146/annurev.astro.46.060407.145222}, \href
  {http://adsabs.harvard.edu/abs/2009ARA%26A..47..481A} {47, 481}

\bibitem[\protect\citeauthoryear{{Bailey}, {Landstreet}  \& {Bagnulo}}{{Bailey}
  et~al.}{2014}]{BaileyBaLaBa2014}
{Bailey} J.~D.,  {Landstreet} J.~D.,   {Bagnulo} S.,  2014, \mn@doi [\aap]
  {10.1051/0004-6361/201322853}, \href
  {http://adsabs.harvard.edu/abs/2014A%26A...561A.147B} {561, A147}

\bibitem[\protect\citeauthoryear{Balmforth, Cunha, Dolez, Gough  \&
  Vauclair}{Balmforth et~al.}{2001}]{BalmforthBaCuDoetal2001}
Balmforth N.~J.,  Cunha M.~S.,  Dolez N.,  Gough D.~O.,   Vauclair S.,  2001,
  \mn@doi [\mnras] {10.1046/j.1365-8711.2001.04182.x}, 323, 362

\bibitem[\protect\citeauthoryear{{Berger}}{{Berger}}{1956}]{BergerJ1956}
{Berger} J.,  1956, Contribution de l'Institut d'Astrophysique de Paris Series
  A, 217

\bibitem[\protect\citeauthoryear{{Briquet}, {Aerts}, {L{\"u}ftinger}, {De Cat},
  {Piskunov}  \& {Scuflaire}}{{Briquet}
  et~al.}{2004}]{BriquetAeLuDeCatetal2004}
{Briquet} M.,  {Aerts} C.,  {L{\"u}ftinger} T.,  {De Cat} P.,  {Piskunov}
  N.~E.,   {Scuflaire} R.,  2004, \mn@doi [\aap] {10.1051/0004-6361:20031450},
  \href {http://adsabs.harvard.edu/abs/2004A%26A...413..273B} {413, 273}

\bibitem[\protect\citeauthoryear{{Bruntt} et~al.,}{{Bruntt}
  et~al.}{2008}]{BrunttBrNoCuetal2008}
{Bruntt} H.,  et~al., 2008, \mn@doi [\mnras]
  {10.1111/j.1365-2966.2008.13167.x}, \href
  {http://adsabs.harvard.edu/abs/2008MNRAS.386.2039B} {386, 2039}

\bibitem[\protect\citeauthoryear{{Carrier, F.}, {North, P.}, {Udry, S.}  \&
  {Babel, J.}}{{Carrier, F.} et~al.}{2002}]{Carrier-F.CaNoUdetal2002}
{Carrier, F.} {North, P.} {Udry, S.}  {Babel, J.} 2002, \mn@doi [A&A]
  {10.1051/0004-6361:20021122}, 394, 151

\bibitem[\protect\citeauthoryear{{Castelli}}{{Castelli}}{1998}]{CastelliCa1998}
{Castelli} F.,  1998, Contributions of the Astronomical Observatory Skalnate
  Pleso, \href {http://adsabs.harvard.edu/abs/1998CoSka..27..192C} {27, 192}

\bibitem[\protect\citeauthoryear{{Castelli}, {Parthasarathy}  \&
  {Hack}}{{Castelli} et~al.}{1997}]{CastelliFPaMandHackM1997}
{Castelli} F.,  {Parthasarathy} M.,   {Hack} M.,  1997, \aap, \href
  {https://ui.adsabs.harvard.edu/abs/1997A%26A...321..254C} {321, 254}

\bibitem[\protect\citeauthoryear{{Castro} et~al.,}{{Castro}
  et~al.}{2017}]{CastroFoHuJaetal2017}
{Castro} N.,  et~al., 2017, \mn@doi [\aap] {10.1051/0004-6361/201629751}, \href
  {http://adsabs.harvard.edu/abs/2017A%26A...597L...6C} {597, L6}

\bibitem[\protect\citeauthoryear{{Catanzaro}, {Frasca}, {Molenda-{\.Z}akowicz}
  \& {Marilli}}{{Catanzaro} et~al.}{2010}]{CatanzaroFrMoMaetal2010}
{Catanzaro} G.,  {Frasca} A.,  {Molenda-{\.Z}akowicz} J.,   {Marilli} E.,
  2010, \mn@doi [\aap] {10.1051/0004-6361/201014189}, \href
  {http://adsabs.harvard.edu/abs/2010A%26A...517A...3C} {517, A3}

\bibitem[\protect\citeauthoryear{{Cowley}, {Hartoog}, {Aller}  \&
  {Cowley}}{{Cowley} et~al.}{1973}]{CowleyCoHaAletal1973}
{Cowley} C.~R.,  {Hartoog} M.~R.,  {Aller} M.~F.,   {Cowley} A.~P.,  1973,
  \mn@doi [\apj] {10.1086/152214}, \href
  {http://adsabs.harvard.edu/abs/1973ApJ...183..127C} {183, 127}

\bibitem[\protect\citeauthoryear{{Cowley}, {Elste}  \& {Urbanski}}{{Cowley}
  et~al.}{1978}]{CowleyCoElUr1978}
{Cowley} C.~R.,  {Elste} G.~H.,   {Urbanski} J.~L.,  1978, \mn@doi [\pasp]
  {10.1086/130379}, \href {http://adsabs.harvard.edu/abs/1978PASP...90..536C}
  {90, 536}

\bibitem[\protect\citeauthoryear{{Cowley}, {Ryabchikova}, {Kupka}, {Bord},
  {Mathys}  \& {Bidelman}}{{Cowley} et~al.}{2000}]{CowleyCoRyKuetal2000}
{Cowley} C.~R.,  {Ryabchikova} T.,  {Kupka} F.,  {Bord} D.~J.,  {Mathys} G.,
  {Bidelman} W.~P.,  2000, \mn@doi [\mnras] {10.1046/j.1365-8711.2000.03578.x},
  \href {http://adsabs.harvard.edu/abs/2000MNRAS.317..299C} {317, 299}

\bibitem[\protect\citeauthoryear{{Cunha}, {Alentiev}, {Brand{\~a}o}  \&
  {Perraut}}{{Cunha} et~al.}{2013}]{CunhaCuAlBretal2013}
{Cunha} M.~S.,  {Alentiev} D.,  {Brand{\~a}o} I.~M.,   {Perraut} K.,  2013,
  \mn@doi [\mnras] {10.1093/mnras/stt1679}, \href
  {http://adsabs.harvard.edu/abs/2013MNRAS.436.1639C} {436, 1639}

\bibitem[\protect\citeauthoryear{{Deutsch}}{{Deutsch}}{1947}]{DeutschAJ1947}
{Deutsch} A.~J.,  1947, \mn@doi [\apj] {10.1086/144904}, \href
  {http://adsabs.harvard.edu/abs/1947ApJ...105..283D} {105, 283}

\bibitem[\protect\citeauthoryear{{Drake}, {Nesvacil}, {Hubrig}, {Kochukhov},
  {de La Reza}, {Polosukhina}  \& {Gonzalez}}{{Drake}
  et~al.}{2005}]{DrakeDrNeHuetal2005}
{Drake} N.~A.,  {Nesvacil} N.,  {Hubrig} S.,  {Kochukhov} O.,  {de La Reza} R.,
   {Polosukhina} N.~S.,   {Gonzalez} J.~F.,  2005, in {Hill} V.,  {Francois}
  P.,   {Primas} F.,  eds,  IAU Symposium Vol. 228, From Lithium to Uranium:
  Elemental Tracers of Early Cosmic Evolution. pp 89--90

\bibitem[\protect\citeauthoryear{{Elkin}, {Kurtz}, {Freyhammer}, {Hubrig}  \&
  {Mathys}}{{Elkin} et~al.}{2008}]{ElkinElKuFretal2008}
{Elkin} V.~G.,  {Kurtz} D.~W.,  {Freyhammer} L.~M.,  {Hubrig} S.,   {Mathys}
  G.,  2008, \mn@doi [\mnras] {10.1111/j.1365-2966.2008.13819.x}, \href
  {http://adsabs.harvard.edu/abs/2008MNRAS.390.1250E} {390, 1250}

\bibitem[\protect\citeauthoryear{{Elkin}, {Kurtz}, {Mathys}  \&
  {Freyhammer}}{{Elkin} et~al.}{2010}]{ElkinElKuMaetal2010}
{Elkin} V.~G.,  {Kurtz} D.~W.,  {Mathys} G.,   {Freyhammer} L.~M.,  2010,
  \mn@doi [\mnras] {10.1111/j.1745-3933.2010.00844.x}, \href
  {http://adsabs.harvard.edu/abs/2010MNRAS.404L.104E} {404, L104}

\bibitem[\protect\citeauthoryear{{Elkin}, {Kurtz}, {Shibahashi}  \&
  {Saio}}{{Elkin} et~al.}{2014}]{ElkinElKuShetal2014}
{Elkin} V.~G.,  {Kurtz} D.~W.,  {Shibahashi} H.,   {Saio} H.,  2014, \mn@doi
  [\mnras] {10.1093/mnras/stu1533}, \href
  {http://adsabs.harvard.edu/abs/2014MNRAS.444.1344E} {444, 1344}

\bibitem[\protect\citeauthoryear{{Engmann} \& {Cousineau}}{{Engmann} \&
  {Cousineau}}{2011}]{EngmannCousineau11}
{Engmann} S.,  {Cousineau} D.,  2011, J. Appl. Quant. Methods, 6, 1

\bibitem[\protect\citeauthoryear{{Feigelson} \& {Babu}}{{Feigelson} \&
  {Babu}}{2012}]{FeigelsonBabu2012}
{Feigelson} E.~D.,  {Babu} G.~J.,  2012, {Modern Statistical Methods for
  Astronomy, Cambridge, UK: Cambridge University Press}

\bibitem[\protect\citeauthoryear{{Fossati}, {Bagnulo}, {Monier}, {Khan},
  {Kochukhov}, {Landstreet}, {Wade}  \& {Weiss}}{{Fossati}
  et~al.}{2007}]{FossatiFoBaMoetal2007}
{Fossati} L.,  {Bagnulo} S.,  {Monier} R.,  {Khan} S.~A.,  {Kochukhov} O.,
  {Landstreet} J.,  {Wade} G.,   {Weiss} W.,  2007, \mn@doi [\aap]
  {10.1051/0004-6361:20078320}, \href
  {http://adsabs.harvard.edu/abs/2007A%26A...476..911F} {476, 911}

\bibitem[\protect\citeauthoryear{{Fossati}, {Folsom}, {Bagnulo}, {Grunhut},
  {Kochukhov}, {Landstreet}, {Paladini}  \& {Wade}}{{Fossati}
  et~al.}{2011}]{FossatiFoBaetal2011}
{Fossati} L.,  {Folsom} C.~P.,  {Bagnulo} S.,  {Grunhut} J.~H.,  {Kochukhov}
  O.,  {Landstreet} J.~D.,  {Paladini} C.,   {Wade} G.~A.,  2011, \mn@doi
  [\mnras] {10.1111/j.1365-2966.2011.18199.x}, \href
  {http://adsabs.harvard.edu/abs/2011MNRAS.413.1132F} {413, 1132}

\bibitem[\protect\citeauthoryear{{Freyhammer}, {Elkin}, {Kurtz}, {Mathys}  \&
  {Martinez}}{{Freyhammer} et~al.}{2008}]{FreyhammerFrElKuetal2008}
{Freyhammer} L.~M.,  {Elkin} V.~G.,  {Kurtz} D.~W.,  {Mathys} G.,   {Martinez}
  P.,  2008, \mn@doi [\mnras] {10.1111/j.1365-2966.2008.13595.x}, \href
  {http://adsabs.harvard.edu/abs/2008MNRAS.389..441F} {389, 441}

\bibitem[\protect\citeauthoryear{{Gelbmann}}{{Gelbmann}}{1998}]{GelbmannGe1998}
{Gelbmann} M.~J.,  1998, Contributions of the Astronomical Observatory Skalnate
  Pleso, \href {http://adsabs.harvard.edu/abs/1998CoSka..27..280G} {27, 280}

\bibitem[\protect\citeauthoryear{{Gelbmann}, {Ryabchikova}, {Weiss},
  {Piskunov}, {Kupka}  \& {Mathys}}{{Gelbmann}
  et~al.}{2000}]{GelbmannGeRyWeetal2000}
{Gelbmann} M.,  {Ryabchikova} T.,  {Weiss} W.~W.,  {Piskunov} N.,  {Kupka} F.,
   {Mathys} G.,  2000, \aap, \href
  {http://adsabs.harvard.edu/abs/2000A%26A...356..200G} {356, 200}

\bibitem[\protect\citeauthoryear{{Gerbaldi}, {Floquet}, {Faraggiana}  \& {van't
  Veer-Menneret}}{{Gerbaldi} et~al.}{1989}]{GerbaldiGeFlFaetal1989}
{Gerbaldi} M.,  {Floquet} M.,  {Faraggiana} R.,   {van't Veer-Menneret} C.,
  1989, \aaps, \href {http://adsabs.harvard.edu/abs/1989A%26AS...81..127G} {81,
  127}

\bibitem[\protect\citeauthoryear{{Ghazaryan} \& {Alecian}}{{Ghazaryan} \&
  {Alecian}}{2016}]{GhazaryanGhAl2016}
{Ghazaryan} S.,  {Alecian} G.,  2016, \mn@doi [\mnras] {10.1093/mnras/stw911},
  \href {http://adsabs.harvard.edu/abs/2016MNRAS.460.1912G} {460, 1912}

\bibitem[\protect\citeauthoryear{{Ghazaryan}, {Alecian}  \&
  {Hakobyan}}{{Ghazaryan} et~al.}{2018}]{GhazaryanSAlandHa2018}
{Ghazaryan} S.,  {Alecian} G.,   {Hakobyan} A.~A.,  2018, \mn@doi [\mnras]
  {10.1093/mnras/sty1912}, \href
  {http://adsabs.harvard.edu/abs/2018MNRAS.480.2953G} {480, 2953}

\bibitem[\protect\citeauthoryear{{Glagolevskij}, {Leushin}, {Chuntonov}  \&
  {Shulyak}}{{Glagolevskij} et~al.}{2006}]{GlagolevskijLeChShetal2006}
{Glagolevskij} Y.~V.,  {Leushin} V.~V.,  {Chuntonov} G.~A.,   {Shulyak} D.,
  2006, \mn@doi [Astronomy Letters] {10.1134/S1063773706010087}, \href
  {http://adsabs.harvard.edu/abs/2006AstL...32...54G} {32, 54}

\bibitem[\protect\citeauthoryear{{Glagolevskij}, {Leushin}  \&
  {Chountonov}}{{Glagolevskij} et~al.}{2007}]{GlagolevskijLeandCh2007}
{Glagolevskij} Y.~V.,  {Leushin} V.~V.,   {Chountonov} G.~A.,  2007, \mn@doi
  [Astrophysical Bulletin] {10.1134/S1990341307040037}, \href
  {http://adsabs.harvard.edu/abs/2007AstBu..62..319G} {62, 319}

\bibitem[\protect\citeauthoryear{{Groote}, {Kaufmann}  \& {Lange}}{{Groote}
  et~al.}{1982}]{GrooteKaLaetal1982}
{Groote} D.,  {Kaufmann} J.~P.,   {Lange} A.,  1982, \aaps, \href
  {http://adsabs.harvard.edu/abs/1982A%26AS...50...77G} {50, 77}

\bibitem[\protect\citeauthoryear{{Hack}, {Castelli}, {Polosukhina}  \&
  {Mavanushenko}}{{Hack} et~al.}{1997a}]{HackHaCaPoetal1997}
{Hack} M.,  {Castelli} F.,  {Polosukhina} N.~S.,   {Mavanushenko} V.~P.,
  1997a, \mn@doi [Astronomical and Astrophysical Transactions]
  {10.1080/10556799708202970}, \href
  {http://adsabs.harvard.edu/abs/1997A%26AT...13..283H} {13, 283}

\bibitem[\protect\citeauthoryear{{Hack}, {Polosukhina}, {Malanushenko}  \&
  {Castelli}}{{Hack} et~al.}{1997b}]{HackHaPoMaetal1997}
{Hack} M.,  {Polosukhina} N.~S.,  {Malanushenko} V.~P.,   {Castelli} F.,
  1997b, \aap, \href {http://adsabs.harvard.edu/abs/1997A%26A...319..637H}
  {319, 637}

\bibitem[\protect\citeauthoryear{{Hubrig}, {Castelli}, {Gonz{\'a}lez}, {Elkin},
  {Mathys}, {Cowley}, {Wolff}  \& {Sch{\"o}ller}}{{Hubrig}
  et~al.}{2012}]{HubrigHuCaGoetal2012}
{Hubrig} S.,  {Castelli} F.,  {Gonz{\'a}lez} J.~F.,  {Elkin} V.~G.,  {Mathys}
  G.,  {Cowley} C.~R.,  {Wolff} B.,   {Sch{\"o}ller} M.,  2012, \mn@doi [\aap]
  {10.1051/0004-6361/201218968}, \href
  {http://adsabs.harvard.edu/abs/2012A%26A...542A..31H} {542, A31}

\bibitem[\protect\citeauthoryear{{Hunger} \& {Groote}}{{Hunger} \&
  {Groote}}{1999}]{HungerandGroote1999}
{Hunger} K.,  {Groote} D.,  1999, \aap, \href
  {http://adsabs.harvard.edu/abs/1999A%26A...351..554H} {351, 554}

\bibitem[\protect\citeauthoryear{{Hunger}, {Groote}  \& {Heber}}{{Hunger}
  et~al.}{1991}]{HungerKGrandHeber1991}
{Hunger} K.,  {Groote} D.,   {Heber} U.,  1991, in {Michaud} G.,  {Tutukov}
  A.~V.,  eds,  IAU Symposium Vol. 145, Evolution of Stars: the Photospheric
  Abundance Connection. p.~173

\bibitem[\protect\citeauthoryear{{Joshi}, {Ryabchikova}, {Kochukhov},
  {Sachkov}, {Tiwari}, {Chakradhari}  \& {Piskunov}}{{Joshi}
  et~al.}{2010}]{JoshiJoRyKoetal2010}
{Joshi} S.,  {Ryabchikova} T.,  {Kochukhov} O.,  {Sachkov} M.,  {Tiwari} S.~K.,
   {Chakradhari} N.~K.,   {Piskunov} N.,  2010, \mn@doi [\mnras]
  {10.1111/j.1365-2966.2009.15725.x}, \href
  {http://adsabs.harvard.edu/abs/2010MNRAS.401.1299J} {401, 1299}

\bibitem[\protect\citeauthoryear{{Kato} \& {Sadakane}}{{Kato} \&
  {Sadakane}}{1999}]{KatoKaSa1999}
{Kato} K.-I.,  {Sadakane} K.,  1999, \mn@doi [\pasj] {10.1093/pasj/51.1.23},
  \href {http://adsabs.harvard.edu/abs/1999PASJ...51...23K} {51, 23}

\bibitem[\protect\citeauthoryear{{K{\i}l{\i}{\c c}o{\u g}lu}, {Monier},
  {Richer}, {Fossati}  \& {Albayrak}}{{K{\i}l{\i}{\c c}o{\u g}lu}
  et~al.}{2016}]{KilicogluMoRietal2016}
{K{\i}l{\i}{\c c}o{\u g}lu} T.,  {Monier} R.,  {Richer} J.,  {Fossati} L.,
  {Albayrak} B.,  2016, \mn@doi [\aj] {10.3847/0004-6256/151/3/49}, \href
  {http://adsabs.harvard.edu/abs/2016AJ....151...49K} {151, 49}

\bibitem[\protect\citeauthoryear{{Kochukhov}}{{Kochukhov}}{2003}]{KochukhovKo2003}
{Kochukhov} O.,  2003, \mn@doi [\aap] {10.1051/0004-6361:20030506}, \href
  {http://adsabs.harvard.edu/abs/2003A%26A...404..669K} {404, 669}

\bibitem[\protect\citeauthoryear{{Kochukhov} \& {Bagnulo}}{{Kochukhov} \&
  {Bagnulo}}{2006}]{KochukhovKoBa2006}
{Kochukhov} O.,  {Bagnulo} S.,  2006, \mn@doi [\aap]
  {10.1051/0004-6361:20054596}, \href
  {http://adsabs.harvard.edu/abs/2006A%26A...450..763K} {450, 763}

\bibitem[\protect\citeauthoryear{{Kochukhov}, {Tsymbal}, {Ryabchikova},
  {Makaganyk}  \& {Bagnulo}}{{Kochukhov}
  et~al.}{2006}]{KochukhovKoTsRyetal2006}
{Kochukhov} O.,  {Tsymbal} V.,  {Ryabchikova} T.,  {Makaganyk} V.,   {Bagnulo}
  S.,  2006, \mn@doi [\aap] {10.1051/0004-6361:20065607}, \href
  {http://adsabs.harvard.edu/abs/2006A%26A...460..831K} {460, 831}

\bibitem[\protect\citeauthoryear{{Kochukhov}, {Ryabchikova}, {Bagnulo}  \& {Lo
  Curto}}{{Kochukhov} et~al.}{2008}]{KochukhovKoRyBaetal2008}
{Kochukhov} O.,  {Ryabchikova} T.,  {Bagnulo} S.,   {Lo Curto} G.,  2008,
  \mn@doi [\aap] {10.1051/0004-6361:20079183}, \href
  {http://adsabs.harvard.edu/abs/2008A%26A...479L..29K} {479, L29}

\bibitem[\protect\citeauthoryear{{Kochukhov}, {Shulyak}  \&
  {Ryabchikova}}{{Kochukhov} et~al.}{2009}]{KochukhovKoShRy2009}
{Kochukhov} O.,  {Shulyak} D.,   {Ryabchikova} T.,  2009, \mn@doi [\aap]
  {10.1051/0004-6361/200911653}, \href
  {http://adsabs.harvard.edu/abs/2009A%26A...499..851K} {499, 851}

\bibitem[\protect\citeauthoryear{Krti{\v c}ka, Kub{\'a}t  \& Groote}{Krti{\v
  c}ka et~al.}{2006}]{KrtickaKrKuGr2006}
Krti{\v c}ka J.,  Kub{\'a}t J.,   Groote D.,  2006, \mn@doi [\aap]
  {10.1051/0004-6361:20065128}, 460, 145

\bibitem[\protect\citeauthoryear{{Krti{\v c}ka}, {Mikul{\'a}{\v s}ek}, {Henry},
  {Zverko}, {{\v Z}i{\v z}ovsk{\'y}}, {Skalick{\'y}}  \& {Zv{\v e}{\v
  r}ina}}{{Krti{\v c}ka} et~al.}{2009}]{KrtickaMiHeetal2009}
{Krti{\v c}ka} J.,  {Mikul{\'a}{\v s}ek} Z.,  {Henry} G.~W.,  {Zverko} J.,
  {{\v Z}i{\v z}ovsk{\'y}} J.,  {Skalick{\'y}} J.,   {Zv{\v e}{\v r}ina} P.,
  2009, \mn@doi [\aap] {10.1051/0004-6361/200811123}, \href
  {http://adsabs.harvard.edu/abs/2009A%26A...499..567K} {499, 567}

\bibitem[\protect\citeauthoryear{{Kupka}, {Ryabchikova}, {Bolgova}, {Kuschnig},
  {Weiss}, {Mathys}  \& {Le Contel}}{{Kupka}
  et~al.}{1994}]{KupkaKuRyBoetal1994}
{Kupka} F.,  {Ryabchikova} T.,  {Bolgova} G.,  {Kuschnig} R.,  {Weiss} W.~W.,
  {Mathys} G.,   {Le Contel} J.~M.,  1994, in {Zverko} J.,  {Ziznovsky} J.,
  eds, Chemically Peculiar and Magnetic Stars. p.~130

\bibitem[\protect\citeauthoryear{{Kupka}, {Paunzen}, {Iliev}  \&
  {Maitzen}}{{Kupka} et~al.}{2004}]{KupkaKuPaIletal2004}
{Kupka} F.,  {Paunzen} E.,  {Iliev} I.~K.,   {Maitzen} H.~M.,  2004, \mn@doi
  [\mnras] {10.1111/j.1365-2966.2004.07977.x}, \href
  {http://adsabs.harvard.edu/abs/2004MNRAS.352..863K} {352, 863}

\bibitem[\protect\citeauthoryear{{Kurtz}}{{Kurtz}}{1978}]{KurtzDW1978}
{Kurtz} D.~W.,  1978, Information Bulletin on Variable Stars, \href
  {http://adsabs.harvard.edu/abs/1978IBVS.1436....1K} {1436}

\bibitem[\protect\citeauthoryear{{Kurtz}}{{Kurtz}}{1982}]{KurtzDW1982}
{Kurtz} D.~W.,  1982, \mn@doi [\mnras] {10.1093/mnras/200.3.807}, \href
  {http://adsabs.harvard.edu/abs/1982MNRAS.200..807K} {200, 807}

\bibitem[\protect\citeauthoryear{{Kurtz}, {Elkin}  \& {Mathys}}{{Kurtz}
  et~al.}{2007}]{KurtzKuElMa2007}
{Kurtz} D.~W.,  {Elkin} V.~G.,   {Mathys} G.,  2007, \mn@doi [\mnras]
  {10.1111/j.1365-2966.2007.12109.x}, \href
  {http://adsabs.harvard.edu/abs/2007MNRAS.380..741K} {380, 741}

\bibitem[\protect\citeauthoryear{{Kurtz} et~al.,}{{Kurtz}
  et~al.}{2011}]{KurtzKuCuSaetal2011}
{Kurtz} D.~W.,  et~al., 2011, \mn@doi [\mnras]
  {10.1111/j.1365-2966.2011.18572.x}, \href
  {http://adsabs.harvard.edu/abs/2011MNRAS.414.2550K} {414, 2550}

\bibitem[\protect\citeauthoryear{{Landstreet}}{{Landstreet}}{1988}]{LandstreetLa1988}
{Landstreet} J.~D.,  1988, \mn@doi [\apj] {10.1086/166155}, \href
  {http://adsabs.harvard.edu/abs/1988ApJ...326..967L} {326, 967}

\bibitem[\protect\citeauthoryear{{LeBlanc}}{{LeBlanc}}{2010}]{LeBlancFr2010}
{LeBlanc} F.,  2010, {An Introduction to Stellar Astrophysics}

\bibitem[\protect\citeauthoryear{{Martinez}, {Kurtz}  \& {Heller}}{{Martinez}
  et~al.}{1990}]{MartinezMaKuHe1990}
{Martinez} P.,  {Kurtz} D.~W.,   {Heller} C.~H.,  1990, \mnras, \href
  {http://adsabs.harvard.edu/abs/1990MNRAS.246..699M} {246, 699}

\bibitem[\protect\citeauthoryear{{Matthews}, {Kurtz}  \& {Martinez}}{{Matthews}
  et~al.}{1999}]{MatthewsMaKuMa1999}
{Matthews} J.~M.,  {Kurtz} D.~W.,   {Martinez} P.,  1999, \mn@doi [\apj]
  {10.1086/306661}, \href {http://adsabs.harvard.edu/abs/1999ApJ...511..422M}
  {511, 422}

\bibitem[\protect\citeauthoryear{{Michaud}}{{Michaud}}{1970}]{MichaudMi1970y}
{Michaud} G.,  1970, \mn@doi [ApJ] {10.1086/150459}, \href
  {http://adsabs.harvard.edu/abs/1970ApJ...160..641M} {160, 641}

\bibitem[\protect\citeauthoryear{{Michaud}, {Alecian}  \& {Richer}}{{Michaud}
  et~al.}{2015}]{MichaudGeAlGeandRicherJ2015}
{Michaud} G.,  {Alecian} G.,   {Richer} J.,  2015, {Atomic Diffusion in Stars,
  Astronomy and Astrophysics Library, Springer International Publishing,
  Switzerland.}

\bibitem[\protect\citeauthoryear{{Mon}, {Hirata}  \& {Sadakane}}{{Mon}
  et~al.}{1981}]{MonHirataandSadakane1981}
{Mon} M.,  {Hirata} R.,   {Sadakane} K.,  1981, \pasj, \href
  {http://adsabs.harvard.edu/abs/1981PASJ...33..413M} {33, 413}

\bibitem[\protect\citeauthoryear{{Nesvacil}, {Shulyak}, {Ryabchikova},
  {Kochukhov}, {Akberov}  \& {Weiss}}{{Nesvacil}
  et~al.}{2013}]{NesvacilNeShRyetal2013}
{Nesvacil} N.,  {Shulyak} D.,  {Ryabchikova} T.~A.,  {Kochukhov} O.,  {Akberov}
  A.,   {Weiss} W.,  2013, \mn@doi [\aap] {10.1051/0004-6361/201220320}, \href
  {http://adsabs.harvard.edu/abs/2013A%26A...552A..28N} {552, A28}

\bibitem[\protect\citeauthoryear{{Netopil}, {Paunzen}, {H{\"u}mmerich}  \&
  {Bernhard}}{{Netopil} et~al.}{2017}]{NetopilNePaHuetal2017}
{Netopil} M.,  {Paunzen} E.,  {H{\"u}mmerich} S.,   {Bernhard} K.,  2017,
  \mn@doi [\mnras] {10.1093/mnras/stx674}, \href
  {http://adsabs.harvard.edu/abs/2017MNRAS.468.2745N} {468, 2745}

\bibitem[\protect\citeauthoryear{{Niemczura} et~al.,}{{Niemczura}
  et~al.}{2015}]{NiemczuraNiMuSmetal2015}
{Niemczura} E.,  et~al., 2015, \mn@doi [\mnras] {10.1093/mnras/stv528}, \href
  {http://adsabs.harvard.edu/abs/2015MNRAS.450.2764N} {450, 2764}

\bibitem[\protect\citeauthoryear{{Norris}}{{Norris}}{1971}]{NorrisJ1971}
{Norris} J.,  1971, \mn@doi [\apjs] {10.1086/190236}, \href
  {http://adsabs.harvard.edu/abs/1971ApJS...23..193N} {23, 193}

\bibitem[\protect\citeauthoryear{{Osmer} \& {Peterson}}{{Osmer} \&
  {Peterson}}{1974}]{OsmerPSandPeterson1974}
{Osmer} P.~S.,  {Peterson} D.~M.,  1974, \mn@doi [\apj] {10.1086/152597}, \href
  {http://adsabs.harvard.edu/abs/1974ApJ...187..117O} {187, 117}

\bibitem[\protect\citeauthoryear{{Polosukhina} et~al.,}{{Polosukhina}
  et~al.}{2004}]{PolosukhinaPoShDretal2004}
{Polosukhina} N.,  et~al., 2004, in {Zverko} J.,  {Ziznovsky} J.,  {Adelman}
  S.~J.,   {Weiss} W.~W.,  eds,  IAU Symposium Vol. 224, The A-Star Puzzle. pp
  665--672

\bibitem[\protect\citeauthoryear{{Przybilla} et~al.,}{{Przybilla}
  et~al.}{2016}]{PrzybillaFoHuNietal2016}
{Przybilla} N.,  et~al., 2016, \mn@doi [\aap] {10.1051/0004-6361/201527646},
  \href {http://adsabs.harvard.edu/abs/2016A%26A...587A...7P} {587, A7}

\bibitem[\protect\citeauthoryear{{Rachkovskaya}, {Lyubimkov}  \&
  {Rostopchin}}{{Rachkovskaya} et~al.}{2006}]{RachkovskayaLyuRoetal2006}
{Rachkovskaya} T.~M.,  {Lyubimkov} L.~S.,   {Rostopchin} S.~I.,  2006, \mn@doi
  [Astronomy Reports] {10.1134/S1063772906020053}, \href
  {http://adsabs.harvard.edu/abs/2006ARep...50..123R} {50, 123}

\bibitem[\protect\citeauthoryear{{Ryabchikova}}{{Ryabchikova}}{1998}]{RyabchikovaRy1998}
{Ryabchikova} T.,  1998, Contributions of the Astronomical Observatory Skalnate
  Pleso, \href {http://adsabs.harvard.edu/abs/1998CoSka..27..319R} {27, 319}

\bibitem[\protect\citeauthoryear{{Ryabchikova} \& {Romanovskaya}}{{Ryabchikova}
  \& {Romanovskaya}}{2017}]{RyabchikovaRyRo2017}
{Ryabchikova} T.~A.,  {Romanovskaya} A.~M.,  2017, \mn@doi [Astronomy Letters]
  {10.1134/S1063773717040065}, \href
  {http://adsabs.harvard.edu/abs/2017AstL...43..252R} {43, 252}

\bibitem[\protect\citeauthoryear{{Ryabchikova}, {Adelman}, {Weiss}  \&
  {Kuschnig}}{{Ryabchikova} et~al.}{1997a}]{RyabchikovaRyAdWeetal1997}
{Ryabchikova} T.~A.,  {Adelman} S.~J.,  {Weiss} W.~W.,   {Kuschnig} R.,  1997a,
  \aap, \href {http://adsabs.harvard.edu/abs/1997A%26A...322..234R} {322, 234}

\bibitem[\protect\citeauthoryear{{Ryabchikova}, {Landstreet}, {Gelbmann},
  {Bolgova}, {Tsymbal}  \& {Weiss}}{{Ryabchikova}
  et~al.}{1997b}]{RyabchikovaRyLaGeetal1997}
{Ryabchikova} T.~A.,  {Landstreet} J.~D.,  {Gelbmann} M.~J.,  {Bolgova} G.~T.,
  {Tsymbal} V.~V.,   {Weiss} W.~W.,  1997b, \aap, \href
  {http://adsabs.harvard.edu/abs/1997A%26A...327.1137R} {327, 1137}

\bibitem[\protect\citeauthoryear{{Ryabchikova}, {Savanov}, {Hatzes}, {Weiss}
  \& {Handler}}{{Ryabchikova} et~al.}{2000}]{RyabchikovaRySaHaetal2000}
{Ryabchikova} T.~A.,  {Savanov} I.~S.,  {Hatzes} A.~P.,  {Weiss} W.~W.,
  {Handler} G.,  2000, \aap, \href
  {http://adsabs.harvard.edu/abs/2000A%26A...357..981R} {357, 981}

\bibitem[\protect\citeauthoryear{{Ryabchikova}, {Savanov}, {Malanushenko}  \&
  {Kudryavtsev}}{{Ryabchikova} et~al.}{2001}]{RyabchikovaRySaMaetal2001}
{Ryabchikova} T.~A.,  {Savanov} I.~S.,  {Malanushenko} V.~P.,   {Kudryavtsev}
  D.~O.,  2001, \mn@doi [Astronomy Reports] {10.1134/1.1369801}, \href
  {http://adsabs.harvard.edu/abs/2001ARep...45..382R} {45, 382}

\bibitem[\protect\citeauthoryear{{Ryabchikova}, {Nesvacil}, {Weiss},
  {Kochukhov}  \& {St{\"u}tz}}{{Ryabchikova}
  et~al.}{2004}]{RyabchikovaRyNeWeetal2004}
{Ryabchikova} T.,  {Nesvacil} N.,  {Weiss} W.~W.,  {Kochukhov} O.,
  {St{\"u}tz} C.,  2004, \mn@doi [\aap] {10.1051/0004-6361:20041012}, \href
  {http://adsabs.harvard.edu/abs/2004A%26A...423..705R} {423, 705}

\bibitem[\protect\citeauthoryear{{Ryabchikova}, {Leone}  \&
  {Kochukhov}}{{Ryabchikova} et~al.}{2005}]{RyabchikovaRyLeKo2005}
{Ryabchikova} T.,  {Leone} F.,   {Kochukhov} O.,  2005, \mn@doi [\aap]
  {10.1051/0004-6361:20041996}, \href
  {http://adsabs.harvard.edu/abs/2005A%26A...438..973R} {438, 973}

\bibitem[\protect\citeauthoryear{{Ryabchikova} et~al.,}{{Ryabchikova}
  et~al.}{2006}]{RyabchikovaRyKoKuetal2006}
{Ryabchikova} T.,  et~al., 2006, \mn@doi [\aap] {10.1051/0004-6361:200500222},
  \href {http://adsabs.harvard.edu/abs/2006A%26A...445L..47R} {445, L47}

\bibitem[\protect\citeauthoryear{{Ryabchikova}, {Kochukhov}  \&
  {Bagnulo}}{{Ryabchikova} et~al.}{2008}]{RyabchikovaRyKoBa2008}
{Ryabchikova} T.,  {Kochukhov} O.,   {Bagnulo} S.,  2008, \mn@doi [\aap]
  {10.1051/0004-6361:20077834}, \href
  {http://adsabs.harvard.edu/abs/2008A%26A...480..811R} {480, 811}

\bibitem[\protect\citeauthoryear{{Sadakane}, {Takada}  \& {Jugaku}}{{Sadakane}
  et~al.}{1983}]{SadakaneSaTaJu1983}
{Sadakane} K.,  {Takada} M.,   {Jugaku} J.,  1983, \mn@doi [\apj]
  {10.1086/161444}, \href {http://adsabs.harvard.edu/abs/1983ApJ...274..261S}
  {274, 261}

\bibitem[\protect\citeauthoryear{{Saffe} \& {Levato}}{{Saffe} \&
  {Levato}}{2014}]{SaffeCandLevatoH2014}
{Saffe} C.,  {Levato} H.,  2014, \mn@doi [\aap] {10.1051/0004-6361/201322091},
  \href {http://adsabs.harvard.edu/abs/2014A%26A...562A.128S} {562, A128}

\bibitem[\protect\citeauthoryear{{Sargent}}{{Sargent}}{1964}]{SargentWLW1964}
{Sargent} W.~L.~W.,  1964, \mn@doi [\araa]
  {10.1146/annurev.aa.02.090164.001501}, \href
  {http://adsabs.harvard.edu/abs/1964ARA%26A...2..297S} {2, 297}

\bibitem[\protect\citeauthoryear{{Savanov} \& {Kochukhov}}{{Savanov} \&
  {Kochukhov}}{1998}]{SavanovSaKo1998}
{Savanov} I.~S.,  {Kochukhov} O.~P.,  1998, Astronomy Letters, \href
  {http://adsabs.harvard.edu/abs/1998AstL...24..516S} {24, 516}

\bibitem[\protect\citeauthoryear{{Semenko}, {Sachkov}, {Ryabchikova},
  {Kudryavtsev}  \& {Piskunov}}{{Semenko}
  et~al.}{2008}]{SemenkoEaSaRyaetal2008}
{Semenko} E.~A.,  {Sachkov} M.~E.,  {Ryabchikova} T.~A.,  {Kudryavtsev} D.~O.,
   {Piskunov} N.~E.,  2008, \mn@doi [Astronomy Letters]
  {10.1134/S1063773708060066}, \href
  {http://adsabs.harvard.edu/abs/2008AstL...34..413S} {34, 413}

\bibitem[\protect\citeauthoryear{{Shavrina} et~al.,}{{Shavrina}
  et~al.}{2001}]{ShavrinaShPoZvetal2001}
{Shavrina} A.~V.,  et~al., 2001, \mn@doi [\aap] {10.1051/0004-6361:20010506},
  \href {http://adsabs.harvard.edu/abs/2001A%26A...372..571S} {372, 571}

\bibitem[\protect\citeauthoryear{{Shavrina} et~al.,}{{Shavrina}
  et~al.}{2004}]{ShavrinaShPoKhetal2004}
{Shavrina} A.,  et~al., 2004, in {Zverko} J.,  {Ziznovsky} J.,  {Adelman}
  S.~J.,   {Weiss} W.~W.,  eds,  IAU Symposium Vol. 224, The A-Star Puzzle. pp
  711--715

\bibitem[\protect\citeauthoryear{{Shavrina}, {Polosukhina}, {Drake},
  {Kudryavtsev}, {Gopka}, {Yushchenko}  \& {Yushchenko}}{{Shavrina}
  et~al.}{2013a}]{ShavrinaShPoDretal2013}
{Shavrina} A.,  {Polosukhina} N.~S.,  {Drake} N.~A.,  {Kudryavtsev} D.~O.,
  {Gopka} V.~F.,  {Yushchenko} V.~A.,   {Yushchenko} A.~V.,  2013a, preprint,
  \href {http://adsabs.harvard.edu/abs/2013arXiv1304.4175S} {} (\mn@eprint
  {arXiv} {1304.4175})

\bibitem[\protect\citeauthoryear{{Shavrina}, {Khalack}, {Glagolevskij},
  {Lyashko}, {Landstreet}, {Leone}, {Polosukhina}  \& {Giarrusso}}{{Shavrina}
  et~al.}{2013b}]{ShavrinaShKhGletal2013}
{Shavrina} A.~V.,  {Khalack} V.,  {Glagolevskij} Y.,  {Lyashko} D.,
  {Landstreet} J.,  {Leone} F.,  {Polosukhina} N.~S.,   {Giarrusso} M.,  2013b,
  Odessa Astronomical Publications, \href
  {http://adsabs.harvard.edu/abs/2013OAP....26..112S} {26, 112}

\bibitem[\protect\citeauthoryear{{Shultz} et~al.,}{{Shultz}
  et~al.}{2015}]{ShultzRiFoWaetal2015}
{Shultz} M.,  et~al., 2015, \mn@doi [\mnras] {10.1093/mnras/stv564}, \href
  {http://adsabs.harvard.edu/abs/2015MNRAS.449.3945S} {449, 3945}

\bibitem[\protect\citeauthoryear{{Shulyak}, {Ryabchikova}, {Mashonkina}  \&
  {Kochukhov}}{{Shulyak} et~al.}{2009}]{ShulyakShRyMaetal2009}
{Shulyak} D.,  {Ryabchikova} T.,  {Mashonkina} L.,   {Kochukhov} O.,  2009,
  \mn@doi [\aap] {10.1051/0004-6361/200911623}, \href
  {http://adsabs.harvard.edu/abs/2009A%26A...499..879S} {499, 879}

\bibitem[\protect\citeauthoryear{{Shulyak}, {Ryabchikova}, {Kildiyarova}  \&
  {Kochukhov}}{{Shulyak} et~al.}{2010}]{ShulyakShRyKietal2010}
{Shulyak} D.,  {Ryabchikova} T.,  {Kildiyarova} R.,   {Kochukhov} O.,  2010,
  \mn@doi [\aap] {10.1051/0004-6361/200913750}, \href
  {http://adsabs.harvard.edu/abs/2010A%26A...520A..88S} {520, A88}

\bibitem[\protect\citeauthoryear{{Smalley} et~al.,}{{Smalley}
  et~al.}{2015}]{SmalleySmNiMuetal2015}
{Smalley} B.,  et~al., 2015, \mn@doi [\mnras] {10.1093/mnras/stv1515}, \href
  {http://adsabs.harvard.edu/abs/2015MNRAS.452.3334S} {452, 3334}

\bibitem[\protect\citeauthoryear{{Smith}}{{Smith}}{1996}]{SmithSm1996}
{Smith} K.~C.,  1996, \mn@doi [\apss] {10.1007/BF02424427}, \href
  {http://adsabs.harvard.edu/abs/1996Ap%26SS.237...77S} {237, 77}

\bibitem[\protect\citeauthoryear{{Smith} \& {Dworetsky}}{{Smith} \&
  {Dworetsky}}{1993}]{SmithSmDw1993r}
{Smith} K.~C.,  {Dworetsky} M.~M.,  1993, A\&A, \href
  {http://adsabs.harvard.edu/abs/1993A%26A...274..335S} {274, 335}

\bibitem[\protect\citeauthoryear{{Spearman}}{{Spearman}}{1904}]{Spearman1904}
{Spearman} C.,  1904, American Journal of Psychology, 15, 72

\bibitem[\protect\citeauthoryear{{Stift} \& {Alecian}}{{Stift} \&
  {Alecian}}{2016}]{StiftStAl2016}
{Stift} M.~J.,  {Alecian} G.,  2016, \mn@doi [MNRAS] {10.1093/mnras/stv2962},
  \href {http://adsabs.harvard.edu/abs/2016MNRAS.457...74S} {457, 74}

\bibitem[\protect\citeauthoryear{{Takada-Hidai} \& {Takeda}}{{Takada-Hidai} \&
  {Takeda}}{1996}]{Takada-HidaiTaTa1996}
{Takada-Hidai} M.,  {Takeda} Y.,  1996, \mn@doi [\pasj]
  {10.1093/pasj/48.5.739}, \href
  {http://adsabs.harvard.edu/abs/1996PASJ...48..739T} {48, 739}

\bibitem[\protect\citeauthoryear{{Takada-Hidai}, {Sadakane}  \&
  {Jugaku}}{{Takada-Hidai} et~al.}{1986}]{Takada-HidaiTaSaJu1986}
{Takada-Hidai} M.,  {Sadakane} K.,   {Jugaku} J.,  1986, \mn@doi [\apj]
  {10.1086/164177}, \href {http://adsabs.harvard.edu/abs/1986ApJ...304..425T}
  {304, 425}

\bibitem[\protect\citeauthoryear{{Vauclair}}{{Vauclair}}{1975}]{VauclairVa1975u}
{Vauclair} S.,  1975, A\&A, \href
  {http://adsabs.harvard.edu/abs/1975A%26A....45..233V} {45, 233}

\bibitem[\protect\citeauthoryear{{Vilhu}, {Tuominen}  \& {Boyarchuk}}{{Vilhu}
  et~al.}{1976}]{VilhuOTuandBo1976}
{Vilhu} O.,  {Tuominen} I.~V.,   {Boyarchuk} A.~A.,  1976, in {Weiss} W.~W.,
  {Jenkner} H.,   {Wood} H.~J.,  eds, IAU Colloq. 32: Physics of Ap Stars.
  p.~563

\bibitem[\protect\citeauthoryear{{Wahlgren} \& {Hubrig}}{{Wahlgren} \&
  {Hubrig}}{2004}]{WahlgrenandHubrig2004}
{Wahlgren} G.~M.,  {Hubrig} S.,  2004, \mn@doi [\aap]
  {10.1051/0004-6361:20034257}, \href
  {http://adsabs.harvard.edu/abs/2004A%26A...418.1073W} {418, 1073}

\bibitem[\protect\citeauthoryear{{Weiss}, {Ryabchikova}, {Kupka}, {Lueftinger},
  {Savanov}  \& {Malanushenko}}{{Weiss} et~al.}{2000}]{WeissWeRyKuetal2000}
{Weiss} W.~W.,  {Ryabchikova} T.~A.,  {Kupka} F.,  {Lueftinger} T.~R.,
  {Savanov} I.~S.,   {Malanushenko} V.~P.,  2000, in {Szabados} L.,  {Kurtz}
  D.,  eds,  Astronomical Society of the Pacific Conference Series Vol. 203,
  IAU Colloq. 176: The Impact of Large-Scale Surveys on Pulsating Star
  Research. pp 487--488

\bibitem[\protect\citeauthoryear{{Zboril} \& {North}}{{Zboril} \&
  {North}}{1998}]{ZborilandNorth1998}
{Zboril} M.,  {North} P.,  1998, Contributions of the Astronomical Observatory
  Skalnate Pleso, \href {http://adsabs.harvard.edu/abs/1998CoSka..27..371Z}
  {27, 371}

\bibitem[\protect\citeauthoryear{{Zboril} \& {North}}{{Zboril} \&
  {North}}{1999}]{ZborilandNorth1999}
{Zboril} M.,  {North} P.,  1999, \aap, \href
  {http://adsabs.harvard.edu/abs/1999A%26A...345..244Z} {345, 244}

\bibitem[\protect\citeauthoryear{{Zboril} \& {North}}{{Zboril} \&
  {North}}{2000}]{ZborilandNorth2000}
{Zboril} M.,  {North} P.,  2000, Contributions of the Astronomical Observatory
  Skalnate Pleso, \href {http://adsabs.harvard.edu/abs/2000CoSka..30...12Z}
  {30, 12}

\bibitem[\protect\citeauthoryear{{Zboril}, {Glagolevskij}  \& {North}}{{Zboril}
  et~al.}{1994}]{ZborilGlandNorth1994}
{Zboril} M.,  {Glagolevskij} Y.~V.,   {North} P.,  1994, in {Zverko} J.,
  {Ziznovsky} J.,  eds, Chemically Peculiar and Magnetic Stars. p.~105

\bibitem[\protect\citeauthoryear{{Zboril}, {North}, {Glagolevskij}  \&
  {Betrix}}{{Zboril} et~al.}{1997}]{ZborilNoGletal1997}
{Zboril} M.,  {North} P.,  {Glagolevskij} Y.~V.,   {Betrix} F.,  1997, \aap,
  \href {http://adsabs.harvard.edu/abs/1997A%26A...324..949Z} {324, 949}

\makeatother
\end{thebibliography}

\cite{*}


\newpage
\onecolumn
\appendix

\section[]{table}
\label{appendix:table}
This table shows the main physical parameters of the 88 CP stars added or updated in our catalog (see Paper~II). The complete data (including other physical parameters and detailed abundances are available as online data).

\begin{itemize}
\item Column 1: the star identification, HD number if available. ({\it{Star id}})

\item Column 2: another usual identification taken from Simbad archive. ({\it{Other id}})

\item Column 3: the CP type: roAp, He-w, He-r. ({\it{CP type}})

\item Column 4: The multiplicity indicator ({\it{Mult.}}) for each star is given by the numbers 1 to 6, which have following meaning:
\begin{itemize}
\item 1: single star,
\item 2:  star in double system,
\item 3:  star in cluster,
\item 4:  spectroscopic binary,
\item 5:  star in multiple system and
\item 6:  eclipsing binary.
\end{itemize}
\item Column 5 to 7: effective temperature (K), gravity, projected rotation velocity (km~s$^{-1}$) if available, using usual symbols.

\item Column 8: References of publications (listed below the table) from which the physical parameters and chemical abundances of those stars are compiled.
\end{itemize}
\bigskip

\defcitealias{AdelmanAdCaCaetal1999}{5}
\defcitealias{AlentievAlKoRyetal2012}{14}
\defcitealias{BrunttBrNoCuetal2008}{23}
\defcitealias{Carrier-F.CaNoUdetal2002}{31}
\defcitealias{CastelliCa1998}{32}
\defcitealias{CowleyCoHaAletal1973}{38}
\defcitealias{CowleyCoElUr1978}{39}
\defcitealias{CowleyCoRyKuetal2000}{40}
\defcitealias{CunhaCuAlBretal2013}{44}
\defcitealias{DrakeDrNeHuetal2005}{47}
\defcitealias{ElkinElKuFretal2008}{52}
\defcitealias{ElkinElKuMaetal2010}{53}
\defcitealias{ElkinElKuShetal2014}{56}
\defcitealias{FossatiFoBaMoetal2007}{62}
\defcitealias{FreyhammerFrElKuetal2008}{67}
\defcitealias{GelbmannGe1998}{70}
\defcitealias{GelbmannGeRyWeetal2000}{71}
\defcitealias{GerbaldiGeFlFaetal1989}{72}
\defcitealias{HackHaPoMaetal1997}{78}
\defcitealias{HackHaCaPoetal1997}{79}
\defcitealias{HubrigHuCaGoetal2012}{81}
\defcitealias{JoshiJoRyKoetal2010}{88}
\defcitealias{KatoKaSa1999}{89}
\defcitealias{KochukhovKo2003}{93}
\defcitealias{KochukhovKoBa2006}{95}
\defcitealias{KochukhovKoTsRyetal2006}{96}
\defcitealias{KochukhovKoRyBaetal2008}{97}
\defcitealias{KochukhovKoShRy2009}{98}
\defcitealias{KupkaKuRyBoetal1994}{100}
\defcitealias{KupkaKuPaIletal2004}{101}
\defcitealias{KurtzKuElMa2007}{102}
\defcitealias{KurtzKuCuSaetal2011}{103}
\defcitealias{LandstreetLa1988}{105}
\defcitealias{MartinezMaKuHe1990}{118}
\defcitealias{MatthewsMaKuMa1999}{119}
\defcitealias{NesvacilNeShRyetal2013}{124}
\defcitealias{NetopilNePaHuetal2017}{126}
\defcitealias{NiemczuraNiMuSmetal2015}{130}
\defcitealias{PolosukhinaPoShDretal2004}{134}
\defcitealias{Allende-PrietoAlLa1999}{135}
\defcitealias{RyabchikovaRyAdWeetal1997}{144}
\defcitealias{RyabchikovaRyLaGeetal1997}{145}
\defcitealias{RyabchikovaRy1998}{146}
\defcitealias{RyabchikovaRySaHaetal2000}{149}
\defcitealias{RyabchikovaRySaMaetal2001}{150}
\defcitealias{RyabchikovaRyNeWeetal2004}{151}
\defcitealias{RyabchikovaRyLeKo2005}{152}
\defcitealias{RyabchikovaRyKoKuetal2006}{153}
\defcitealias{RyabchikovaRyKoBa2008}{155}
\defcitealias{RyabchikovaRyRo2017}{156}
\defcitealias{SadakaneSaTaJu1983}{157}
\defcitealias{SavanovSaKo1998}{160}
\defcitealias{SemenkoEaSaRyaetal2008}{162}
\defcitealias{ShavrinaShPoZvetal2001}{164}
\defcitealias{ShavrinaShPoKhetal2004}{165}
\defcitealias{ShavrinaShKhGletal2013}{166}
\defcitealias{ShavrinaShPoDretal2013}{167}
\defcitealias{ShulyakShRyMaetal2009}{169}
\defcitealias{ShulyakShRyKietal2010}{170}
\defcitealias{SmalleySmNiMuetal2015}{172}
\defcitealias{Takada-HidaiTaSaJu1986}{182}
\defcitealias{Takada-HidaiTaTa1996}{183}
\defcitealias{WeissWeRyKuetal2000}{195}
\defcitealias{KilicogluMoRietal2016}{205}
\defcitealias{ShultzRiFoWaetal2015}{206}
\defcitealias{SaffeCandLevatoH2014}{207}
\defcitealias{FossatiFoBaetal2011}{209}
\defcitealias{CatanzaroFrMoMaetal2010}{210}
\defcitealias{KrtickaMiHeetal2009}{211}
\defcitealias{GlagolevskijLeandCh2007}{212}
\defcitealias{RachkovskayaLyuRoetal2006}{213}
\defcitealias{GlagolevskijLeChShetal2006}{214}
\defcitealias{WahlgrenandHubrig2004}{215}
\defcitealias{BriquetAeLuDeCatetal2004}{216}
\defcitealias{CastelliFPaMandHackM1997}{217}
\defcitealias{MonHirataandSadakane1981}{218}
\defcitealias{CastroFoHuJaetal2017}{219}
\defcitealias{PrzybillaFoHuNietal2016}{220}
\defcitealias{ZborilandNorth2000}{221}
\defcitealias{ZborilandNorth1999}{222}
\defcitealias{HungerandGroote1999}{223}
\defcitealias{ZborilandNorth1998}{224}
\defcitealias{ZborilNoGletal1997}{225}
\defcitealias{ZborilGlandNorth1994}{226}
\defcitealias{GrooteKaLaetal1982}{227}

\tablefirsthead{%
\hline
\multicolumn{1}{|l|}{\textbf{Star id}} &
\multicolumn{1}{c|}{\textbf{Other id}} &
\multicolumn{1}{c|}{\textbf{CP type}} &
\multicolumn{1}{c|}{\textbf{Mult.}} &
\multicolumn{1}{c|}{$T_{\rm{eff}}$\,(K)} &
\multicolumn{1}{c|}{$\log{g}$} &
\multicolumn{1}{c|}{$v\,\sin{i}$} &
\multicolumn{1}{c|}{\textbf{References}}\\
\hline }

\tablehead{%
\hline
\multicolumn{8}{|c|}{\small\sl continued from previous page}\\
\hline
\multicolumn{1}{|l|}{\textbf{Star id}} &
\multicolumn{1}{c|}{\textbf{Other id}} &
\multicolumn{1}{c|}{\textbf{CP type}} &
\multicolumn{1}{c|}{\textbf{Mult.}} &
\multicolumn{1}{c|}{$T_{\rm{eff}}$\,(K)} &
\multicolumn{1}{c|}{$\log{g}$} &
\multicolumn{1}{c|}{$v\,\sin{i}$} &
\multicolumn{1}{c|}{\textbf{References}}\\
\hline }
\tabletail{%
   \hline}
\tablelasttail{%
   \hline
   \multicolumn{8}{|c|}{\small\sl table end}\\
   \hline}

\centering
\begin{supertabular}{|l|c|c|c|c|c|c|c|}
\topcaption{CP stars and their fundamental parameters}
\label{table:fund}
HD\,201601&$\gamma\,Equ$&roAp&2&7750&4.2&4.5&\citetalias{GelbmannGe1998},\citetalias{HubrigHuCaGoetal2012},\citetalias{RyabchikovaRyAdWeetal1997},\citetalias{ShavrinaShPoKhetal2004},\citetalias{Takada-HidaiTaTa1996}\\
HD\,203932&$BI\,Mic$&roAp&1&7450&4.3&12.5&\citetalias{GelbmannGe1998},\citetalias{RyabchikovaRySaHaetal2000}\\
HD\,24712&$DO\,Eri$&roAp&1&7250&4.3&7.0&\citetalias{CowleyCoRyKuetal2000},\citetalias{GelbmannGe1998},\citetalias{ShulyakShRyMaetal2009}\\
HD\,217522&HIP\,113711&roAp&1&6750&4.3&12.0&\citetalias{GelbmannGe1998}\\
HD\,166473&TYC\,7900-2776-1&roAp&1&8000&4.4&18.0&\citetalias{GelbmannGe1998},\citetalias{ShavrinaShKhGletal2013}\\
HD\,128898&$\alpha\,Cir$&roAp&2&7900&4.2&12.5&\citetalias{BrunttBrNoCuetal2008},\citetalias{GelbmannGe1998},\citetalias{KochukhovKoShRy2009}\\
HD\,204411&HR\,8216&roAp&3&8400&3.5&5.4&\citetalias{CowleyCoElUr1978},\citetalias{KochukhovKoTsRyetal2006},\citetalias{RyabchikovaRyLeKo2005}\\
HD\,69013&TYC\,5996-1937-1&roAp&1&7500&4.5&4.0&\citetalias{FreyhammerFrElKuetal2008},\citetalias{RyabchikovaRyRo2017}\\
HD\,96237&TYC\,6640-1026-1&roAp&1&7800&4.43&6.0&\citetalias{FreyhammerFrElKuetal2008},\citetalias{RyabchikovaRyRo2017}\\
HD\,118022&HR\,5105&roAp&1&9950&4&10.0&\citetalias{GerbaldiGeFlFaetal1989},\citetalias{RyabchikovaRyRo2017}\\
HD\,188041&HR\,7575&roAp&1&8800&4&4.0&\citetalias{GerbaldiGeFlFaetal1989},\citetalias{KatoKaSa1999},\citetalias{RyabchikovaRyNeWeetal2004},\citetalias{RyabchikovaRyRo2017}\\
TYC\,3545-2756-1&KIC\,10195926&roAp&1&7200&3.6&21.0&\citetalias{ElkinElKuShetal2014},\citetalias{KurtzKuCuSaetal2011}\\
HD\,101065&HIP\,56709&roAp&1&6622&4.06&3.5&\citetalias{CowleyCoRyKuetal2000},\citetalias{RyabchikovaRyKoBa2008},\citetalias{ShavrinaShPoKhetal2004},\citetalias{ShulyakShRyKietal2010}\\
HD\,103498&HR\,4561&roAp&2&9500&3.6&12.0&\citetalias{JoshiJoRyKoetal2010}\\
HD\,115226&HIP\,64883&roAp&1&7640&4&27.5&\citetalias{KochukhovKoRyBaetal2008}\\
HD\,178892&HIP\,94155&roAp&2&7700&4&9.0&\citetalias{RyabchikovaRyKoKuetal2006}\\
HD\,92499&HIP\,52218&roAp&1&7810&4&2.0&\citetalias{ElkinElKuMaetal2010},\citetalias{FreyhammerFrElKuetal2008}\\
HD\,143487&TYC\,7329-1814-1&roAp&4&6930&4&2.0&\citetalias{ElkinElKuMaetal2010},\citetalias{FreyhammerFrElKuetal2008}\\
HD\,65339&53\,Cam&roAp&1&8500&4&13.0&\citetalias{GerbaldiGeFlFaetal1989},\citetalias{LandstreetLa1988}\\
HD\,154708&HIP\,84017&roAp&4&6800&4.11&6.0&\citetalias{HubrigHuCaGoetal2012}\\
HD\,184471&HIP\,96177&roAp&1&7500&4&10.0&\citetalias{RyabchikovaRySaMaetal2001},\citetalias{WeissWeRyKuetal2000}\\
HD\,42659&HIP\,29635&roAp&4&8100&4.2&19.0&\citetalias{RyabchikovaRySaMaetal2001},\citetalias{WeissWeRyKuetal2000}\\
HD\,176232&10\,Aql&roAp&1&7550&4&4.0&\citetalias{CunhaCuAlBretal2013},\citetalias{NesvacilNeShRyetal2013},\citetalias{RyabchikovaRySaHaetal2000},\citetalias{RyabchikovaRyNeWeetal2004},\citetalias{SavanovSaKo1998}\\
HD\,122970&HIP\,68790&roAp&1&6930&4.4&5.5&\citetalias{CowleyCoRyKuetal2000},\citetalias{KochukhovKo2003},\citetalias{RyabchikovaRySaHaetal2000},\citetalias{RyabchikovaRyKoBa2008},\citetalias{WeissWeRyKuetal2000}\\
HD\,137949&33\,Lib&roAp&1&7750&4.5&2.0&\citetalias{RyabchikovaRyNeWeetal2004},\citetalias{ShavrinaShPoKhetal2004}\\
HD\,12098&Renson\,3085&roAp&2&7800&4.3&10.0&\citetalias{RyabchikovaRyNeWeetal2004},\citetalias{ShavrinaShPoDretal2013}\\
HD\,60435&HIP\,36537&roAp&2&8100&4.2&12.0&\citetalias{PolosukhinaPoShDretal2004},\citetalias{RyabchikovaRyNeWeetal2004},\citetalias{ShavrinaShPoZvetal2001}\\
HD\,75445&HIP\,43257&roAp&1&7700&4.3&2.0&\citetalias{RyabchikovaRyNeWeetal2004}\\
HD\,116114&HIP\,65203&roAp&1&8000&4.1&3.0&\citetalias{KurtzKuElMa2007},\citetalias{RyabchikovaRyNeWeetal2004}\\
HD\,137909&$\beta\,CrB$&roAp&1&8000&4.3&3.0&\citetalias{CastelliCa1998},\citetalias{HackHaPoMaetal1997},\citetalias{HackHaCaPoetal1997},\citetalias{KupkaKuRyBoetal1994},\citetalias{RyabchikovaRyNeWeetal2004},\citetalias{Takada-HidaiTaTa1996}\\
HD\,110066&HR\,4816&roAp&1&9000&4.3&9.0&\citetalias{RyabchikovaRy1998},\citetalias{RyabchikovaRyNeWeetal2004}\\
HD\,225914&KIC\,4768731&roAp&1&8100&4&15.0&\citetalias{NiemczuraNiMuSmetal2015},\citetalias{SmalleySmNiMuetal2015}\\
HD\,177765&TYC\,6882-1808-1&roAp&2&8000&3.8&2.5&\citetalias{AlentievAlKoRyetal2012}\\
HD\,965&TYC\,4664-318-1&roAp&1&7500&4&3.0&\citetalias{RyabchikovaRyKoBa2008},\citetalias{ShavrinaShPoDretal2013}\\
HD\,134214&TYC\,5592-971-1&roAp&1&7315&4.45&2.0&\citetalias{RyabchikovaRyKoBa2008},\citetalias{ShavrinaShPoKhetal2004}\\
HD\,213637&TYC\,6391-745-1&roAp&4&6400&3.6&3.5&\citetalias{KochukhovKo2003}\\
HD\,3980&TYC\,8469-1595-1&roAp&1&8100&4&16.5&\citetalias{DrakeDrNeHuetal2005},\citetalias{ElkinElKuFretal2008}\\
HD\,9996a&HR\,465a&roAp&4&10300&3.7&2.0&\citetalias{CowleyCoHaAletal1973},\citetalias{CowleyCoElUr1978}\\
HD\,115708&TYC\,1996-1624-1&roAp&1&7550&4&11.0&\citetalias{SemenkoEaSaRyaetal2008}\\
HD\,83368&HIP\,47145&roAp&2&7750&4&35.0&\citetalias{PolosukhinaPoShDretal2004},\citetalias{ShavrinaShPoZvetal2001}\\
HD\,318101&TYC\,7380-116-1&He\_w&3&15400&4&31.0&\citetalias{KilicogluMoRietal2016}\\
HD\,61556&HR\,2949&He\_w&2&18500&4.1&61.0&\citetalias{ShultzRiFoWaetal2015}\\
HD\,5737&$\alpha\,Scl$&He\_w&1&13600&3.2&17.0&\citetalias{Takada-HidaiTaSaJu1986},\citetalias{SaffeCandLevatoH2014}\\
HD\,123182&TYC\,8268-3344-1&He\_w&3&10800&4.25&81.0&\citetalias{FossatiFoBaetal2011}\\
TYC\,3128-248-1&HIP\,93941&He\_w&1&19300&3.8&10.0&\citetalias{CatanzaroFrMoMaetal2010}\\
HD\,177410&HR\,7224&He\_w&1&14500&4.2&\-&\citetalias{KrtickaMiHeetal2009}\\
HD\,37058&TYC\,4774-927-1&He\_w&3&17000&3.8&25.0&\citetalias{GlagolevskijLeandCh2007}\\
HD\,212454&HR\,8535&He\_w&1&14400&3.83&40.0&\citetalias{SadakaneSaTaJu1983},\citetalias{Takada-HidaiTaSaJu1986},\citetalias{GlagolevskijLeandCh2007}\\
HD\,224926&29\,Psc&He\_w&1&14000&3.85&68.0&\citetalias{GlagolevskijLeandCh2007}\\
HD\,30122&HR\,1512&He\_w&1&15200&3.52&30.0&\citetalias{RachkovskayaLyuRoetal2006}\\
HD\,21699&HR\,1063&He\_w&3&16000&4.15&35.0&\citetalias{GlagolevskijLeChShetal2006}\\
HD\,217833&HR\,8770&He\_w&2&15450&3.88&35.0&\citetalias{SadakaneSaTaJu1983},\citetalias{Takada-HidaiTaSaJu1986},\citetalias{GlagolevskijLeChShetal2006}\\
HD\,120709&3\,Cen\,A&He\_w&2&17500&3.8&20.0&\citetalias{WahlgrenandHubrig2004},\citetalias{CastelliFPaMandHackM1997}\\
HD\,131120&HR\,5543&He\_w&2\,or\,5&18250&4.1&\-&\citetalias{BriquetAeLuDeCatetal2004}\\
HD\,105382&HR\,4618&He\_w&2&17400&4.18&\-&\citetalias{BriquetAeLuDeCatetal2004}\\
HD\,138769&d\,Lup&He\_w&2&17500&4.22&\-&\citetalias{BriquetAeLuDeCatetal2004}\\
HD\,37752&HR\,1951&He\_w&1&15700&3.9&\-&\citetalias{SadakaneSaTaJu1983},\citetalias{Takada-HidaiTaSaJu1986}\\
HD\,44953&HR\,2306&He\_w&2&17000&4.1&\-&\citetalias{SadakaneSaTaJu1983},\citetalias{Takada-HidaiTaSaJu1986}\\
HD\,23408&$20\,Mon$&He\_w&3&12600&3.2&45.0&\citetalias{MonHirataandSadakane1981}\\
HD\,149363&TYC\,5060-966-1&He\_w&1&30000&4&95.0&\citetalias{ZborilandNorth2000}\\
TYC\,8976-5133-1&CPD-622124&He\_r&3&23650&3.95&75.0&\citetalias{CastroFoHuJaetal2017}\\
TYC\,8613-98-1&CPD-573509&He\_r&3&23750&4.05&35.0&\citetalias{PrzybillaFoHuNietal2016}\\
HD\,92938&HR\,4196&He\_r&3&16000&4&125.0&\citetalias{ZborilandNorth2000},\citetalias{ZborilandNorth1999},\citetalias{ZborilGlandNorth1994}\\
HD\,96446&TYC\,8627-156-1&He\_r&1&23000&4&0.0&\citetalias{ZborilandNorth2000},\citetalias{ZborilandNorth1999},\citetalias{ZborilGlandNorth1994}\\
HD\,133518&TYC\,8305-2754-1&He\_r&1&20000&4&0.0&\citetalias{ZborilandNorth2000},\citetalias{ZborilandNorth1999},\citetalias{ZborilGlandNorth1994}\\
HD\,145939&DM-134383&He\_r&1&18000&4&55.0&\citetalias{ZborilandNorth2000}\\
HD\,149257&TYC\,8325-1455-1&He\_r&3&25000&3&40.0&\citetalias{ZborilandNorth2000},\citetalias{ZborilGlandNorth1994}\\
TYC\,9280-38-1&DM-692698&He\_r&1&25000&4&30.0&\citetalias{ZborilandNorth2000},\citetalias{ZborilGlandNorth1994}\\
HD\,164769&TYC\,6850-1755-1&He\_r&1&23000&4&105.0&\citetalias{ZborilandNorth2000}\\
HD\,168785&TYC\,7393-979-1&He\_r&1&23000&4&14.0&\citetalias{ZborilandNorth2000},\citetalias{ZborilGlandNorth1994}\\
HD\,186205&TYC\,1057-69-1&He\_r&1&17000&4&5.0&\citetalias{ZborilandNorth2000}\\
TYC\,7972-666-1&DM-4314300&He\_r&1&22000&4&2.0&\citetalias{ZborilandNorth2000}\\
HD\,36485&HR\,1851&He\_r&4&18400&4.41&54.0&\citetalias{ZborilandNorth1999},\citetalias{ZborilandNorth1998},\citetalias{ZborilNoGletal1997}\\
HD\,37017&HR\,1890&He\_r&4&19200&4.45&80.0&\citetalias{ZborilandNorth1999},\citetalias{ZborilandNorth1998},\citetalias{ZborilNoGletal1997}\\
HD\,37479&HR\,1932&He\_r&3&22200&4.53&100.0&\citetalias{ZborilandNorth1999},\citetalias{ZborilandNorth1998},\citetalias{ZborilNoGletal1997}\\
HD\,37776&HR\,26742&He\_r&1&21800&4.52&75.0&\citetalias{ZborilandNorth1999},\citetalias{ZborilandNorth1998},\citetalias{ZborilNoGletal1997}\\
HD\,260858&TYC\,741-818-1&He\_r&1&19200&4.22&47.0&\citetalias{ZborilandNorth1999},\citetalias{ZborilandNorth1998},\citetalias{ZborilNoGletal1997}\\
HD\,264111&TYC\,156-966-1&He\_r&1&23200&4.54&75.0&\citetalias{ZborilandNorth1999},\citetalias{ZborilandNorth1998},\citetalias{ZborilNoGletal1997}\\
TYC\,6532-2200-1&HIP\,34781&He\_r&1&22700&4.53&45.0&\citetalias{ZborilandNorth1999},\citetalias{ZborilandNorth1998},\citetalias{ZborilNoGletal1997}\\
HD\,58260&HIP\,35830&He\_r&3&19000&4.02&45.0&\citetalias{ZborilandNorth1999},\citetalias{ZborilandNorth1998},\citetalias{ZborilNoGletal1997}\\
HD\,60344&HIP\,36707&He\_r&1&21700&4.48&55.0&\citetalias{ZborilandNorth1999},\citetalias{ZborilandNorth1998},\citetalias{ZborilNoGletal1997}\\
HD\,64740&HR\,3089&He\_r&1&22700&4.50&130.0&\citetalias{ZborilandNorth1999},\citetalias{ZborilandNorth1998},\citetalias{ZborilNoGletal1997}\\
HD\,66522&HIP\,39246&He\_r&1&18800&4.39&0.0&\citetalias{ZborilandNorth1999},\citetalias{ZborilNoGletal1997},\citetalias{ZborilGlandNorth1994}\\
TYC\,8152-1868-1&LS\,1169&He\_r&1&22000&4.52&80.0&\citetalias{ZborilandNorth1999},\citetalias{ZborilNoGletal1997},\citetalias{GrooteKaLaetal1982}\\
HD\,108483&HR\,4743&He\_r&2\,or\,5&19100&4.25&130.0&\citetalias{ZborilandNorth1999},\citetalias{ZborilandNorth1998},\citetalias{ZborilNoGletal1997}\\
HD\,1224448&TYC\,8264-3162-1&He\_r&1&22000&3.70&\-&\citetalias{ZborilNoGletal1997}\\
HD\,169467&$\alpha\,Tel$&He\_r&1&16600&4.05&\-&\citetalias{ZborilGlandNorth1994}\\
HD\,68450&HR\,3219&He\_r&3&32100&3.55&\-&\citetalias{ZborilGlandNorth1994}\\

\end{supertabular}

\textbf{References}: 14- Alentiev et al (2012), 23- Bruntt et al (2008), 32- Castelli (1998), 38- Cowley et al (1973), 39- Cowley et al (1978), 40- Cowley et al (2000), 44- Cunha et al (2013), 47- Drake et al (2005), 52- Elkin et al (2008), 53- Elkin et al (2010), 56- Elkin et al (2014), 67- Freyhammer et al (2008), 70- Gelbmann (1998), 72- Gerbaldi et al (1989), 78- Hack et al (1997a), 79- Hack et al (1997b), 81- Hubrig et al (2012), 88- Joshi et al (2010), 89- Kato and Sadakane (1999), 93- Kochukhov (2003), 96- Kochukhov et al (2006), 97- Kochukhov et al (2008), 98- Kochukhov et al (2009), 100- Kupka et al (1994), 102- Kurtz et al (2007), 103- Kurtz et al (2011), 105- Landstreet (1988), 124- Nesvacil et al (2013), 130- Niemczura et al (2015), 134- Polosukhina et al (2004), 144- Ryabchikova et al (1997a), 146- Ryabchikova (1998), 149- Ryabchikova et al (2000), 150- Ryabchikova et al (2001), 151- Ryabchikova et al (2004a), 152- Ryabchikova et al (2005), 153- Ryabchikova et al (2006), 155- Ryabchikova et al (2008), 156- Ryabchikova and Romanovskaya (2017), 157- Sadakane et al (1983), 160- Savanov and Kochukhov (1998), 164- Shavrina et al (2001), 165- Shavrina et al (2004), 166- Shavrina et al (2013), 167- Shavrina et al (2013a), 169- Shulyak et al (2009), 170- Shulyak et al (2010), 172- Smalley et al (2015), 182- Takada-Hidai et al (1986), 183- Takada-Hidai and Takeda (1996), 195- Weiss et al (2000), 206- Shultz et al (2015), 207- Saffe and Levato (2014), 209- Fossati et al (2011), 210- Catanzaro et al (2010), 211- Krticka et al (2009), 212- Glagolevskij et al (2007), 213- Rachkovskaya et al (2006), 214- Glagolevskij et al (2006), 215- Wahlgren and Hubrig (2004), 216- Briquet et al (2004), 217- Castelli et al (1997), 218- Mon and Hirata (1981), 219- Castro et al (2017), 220- Przybilla et al (2016), 221- Zboril and North (2000), 222- Zboril and North (1999), 224- Zboril and North (1998), 225- Zboril et al (1997), 226- Zboril et al (1994), 227- Groote et al (1982).


\bsp	
\label{lastpage}
\end{document}